\documentclass[twocolumn]{aastex631}

\usepackage{subfigure}
\usepackage{amsmath}
\usepackage{xcolor}
\usepackage{txfonts}

\usepackage{tablefootnote}
\usepackage{pifont}
\usepackage{booktabs}

\newcommand{\psrb}{PSR\,B1259-63/LS\,2883}

\newcommand{\gammaray}{$\gamma$-ray}

\newcommand{\gaga}{$\gamma\gamma$}

\newcommand{\add}[1]{{\color{black} #1}}

\newcommand{\bvs}[1]{{\color{black} #1}}

\newcommand{\is}[1]{{\color{black} #1}}
\newcommand{\isnew}[1]{{\color{black} #1}}
\newcommand{\rev}[1]{{#1}}
\newcommand{\revc}[1]{{#1}}

\begin{document}

\title{Probing orbital parameters of gamma-ray binaries with TeV light curves}

\author[0000-0002-2814-1257]{Iurii Sushch}
\affiliation{Centre for Space Research, North-West University, 2520 Potchefstroom, South Africa}
\affiliation{Astronomical Observatory of Ivan Franko National University of Lviv, Kyryla i Methodia 8, 79005 Lviv, Ukraine}
\email{iurii.sushch@nwu.ac.za}

\author[0000-0003-1873-7855]{Brian van Soelen}
\affiliation{Department of Physics, University of the Free State,  PO Box 339, Bloemfontein 9300, South Africa}
\email{vansoelenb@ufs.ac.za}



\begin{abstract}

Gamma-ray binaries are binary systems where the energy flux peaks in the gamma-ray energy band. They harbour a compact object (a neutron star or a black hole) orbiting around a massive star that provides a strong radiation field. It is believed that the gamma-ray emission from such objects can be strongly attenuated through the electron-positron pair production in gamma-gamma interactions. The importance of gamma-gamma absorption depends on the orbital phase and on the geometry of the system. In this work we propose a method of how the orbital parameters of gamma-ray binaries could be probed with TeV light curves that have imprinted features of gamma-gamma absorption.

\end{abstract}



\section{Introduction} \label{sec:intro}

Gamma-ray binaries constitute a small but growing class of high mass binary systems with spectral energy distributions which  peak above 1 MeV in a $\nu F_\nu$ representation \citep[see][for a detailed review of these systems]{2013A&ARv..21...64D, 2015CRPhy..16..661D,cta_binary19, 2020mbhe.confE..45C}. The class comprises only nine known objects: eight in our Galaxy and one in the Large Magellanic Cloud. These systems consist of a compact source (pulsar or black hole) orbiting around a massive companion star and for only two of them is the nature of the compact source well established -- both PSR\,B1259-63/LS\,2883 and PSR J2032+4127/MT91\,213 comprise of a pulsar. More recently, radio pulsations have been reported for LS\,I~+61\,303 \citet{2022NatAs...6..698W}.    While PSR\,B1259-63/LS\,2883 is a very well studied binary, PSR J2032+4127/MT91\,213 was confirmed as a gamma binary more recently, due to its very long orbital period of $45-50$ years. Very-high-energy gamma-ray emission was detected in 2017 as it passed periastron \citep{2018ApJ...867L..19A}. 

Gamma-ray emission in such plerionic binaries is believed to be produced through the inverse Compton scattering of electrons,  accelerated at the termination shock that forms between the pulsar and stellar winds, on the stellar radiation field and thus is strongly orbital dependant. Additionally, hydrodynamic simulations imply the formation of the Coriolis\footnote{Note, that \citet{2015A&A...581A..27D} argue that the existence of the ``back shock" is not related to Coriolis forces as it also forms in the absence of rotation} shock and other secondary shocks \citep[see e.g.][]{2015A&A...577A..89B, 2015A&A...581A..27D, 2021A&A...646A..91H} resulting in a full confinement of particles and providing additional acceleration and re-acceleration sites. Simulations suggest the highest density of very-high-energy electrons and hence the strongest gamma-ray emission at the apex of the shock \citep{2015A&A...581A..27D, 2021A&A...646A..91H} but this result relies on certain assumptions, e.g. prescription for the particle injection which could be oversimplified. Therefore, although the most plausible location for the emission region is the apex of the shock, emission can be produced at the other sites, located also behind the pulsar with respect to the star, and it is not trivial to address the significance of their contribution.

For the accretion powered systems with a black hole as a compact object (known as microquasars) particles can be accelerated anywhere along the jet through various mechanisms \citep[see e.g.][]{2006A&A...456..801D, 2008MNRAS.383..467K, 2009IJMPD..18..347B}. Recent H.E.S.S. observations of the SS 433 system provide for the first time a direct measurement of the location of the particle acceleration site in a jet-like source which appears to be at a considerable distance of $\sim25$~pc from the central engine \citep{Laura:ss433}. Therefore, the gamma-ray emission in these systems can be produced far off the orbital plane and far from the compact object and its companion star.

Gamma-ray photons emitted in binary systems are subject to $\gamma\gamma$ absorption as they travel through the photon field created by the massive star \citep[see e.g.][]{2006A&A...451....9D}. Moreover, \gaga\ absorption might be the main reason for the characteristic minimum in the TeV light curve observed in gamma-ray binaries \citep{2006A&A...451....9D, 2017ApJ...837..175S}. In the case of PSR\,B1259-63/LS\,2883, \is{a very eccentric binary ($e=0.87$) with an orbital period of 3.4 years,} the TeV flux from the system increases as the pulsar moves closer to the star but then suddenly drops \is{as it gets closer to} periastron itself  and then increases again after the periastron passage forming the second peak\footnote{\isnew{It should be noted that there is some evidence for a less prominent third "middle" peak right before the periastron passage \citep[see the analysis of the H.E.S.S. data for all periastra passages since 2004 in][]{2020A&A...633A.102H}}} \citep{2005A&A...442....1A, 2009A&A...507..389A, 2013A&A...551A..94H, 2015arXiv150903090R, 2020A&A...633A.102H}. This behavior is counter-intuitive, because at periastron the pulsar is located at the closest distance to the star and thus encounters the highest density of the target photon field for inverse Compton radiation. However, these conditions are also optimal for gamma-gamma absorption which might cause a severe decrease of the TeV flux. It was shown for PSR\,B1259-63/LS\,2883 that the orbital phase for which the absorption would be the strongest coincides with the orbital phase of the dip in the light curve \citep{2017ApJ...837..175S} hinting that this decrease of the TeV flux \is{indeed could be related} to the gamma-gamma absorption. \is{The level of absorption is, however, not sufficient to explain the magnitude of the dip in the TeV light curve if the emitting region is located close to the pulsar.} Nevertheless, it is interesting to note that in the phase-folded stacking analysis presented in \citet{2020A&A...633A.102H} there is a hint of a hardening of the TeV spectrum before the periastron ($\sim-20$\,d) which is coincident with a dip in the light curve. This is what would be expected from a change induced by significant gamma-gamma absorption as proposed in \citet{2017ApJ...837..175S}.
\is{For LS~5039, a much more compact binary with low eccentricity and an orbital period of only 3.9 days, the modulation of the light curve could be well explained by gamma-gamma absorption with the opacity at the dip high enough to explain the level of flux attenuation \citep{2006A&A...451....9D,2005ApJ...634L..81B,2021A&A...649A..71H}.} 

\is{If the emitting region is located close enough to the star, i.e. the gamma-gamma opacity is high enough to modulate the TeV light curve,}
 the location of the minimum in the light curve \is{would} contain information on the geometry of the system\is{, because such orbital parameters as  the inclination angle and the  longitude of periastron 
 determine the orbital phase where the absorption would be the highest.}
 

\add{In this paper we show how, if it is assumed that gamma-gamma absorption is significant enough to attenuate the very high energy emission, the location of the minimum in the light curve can be used to place constraints on the orbital geometries the gamma-ray binaries. This is applied to the system PSR J2032+4127/MT91 213.}

\section{Method} \label{sec:method}

\subsection{Gamma-gamma absorption}

The interaction of a \gammaray\ photon of energy $\epsilon_\gamma$ with a low-energy photon of energy $\epsilon$, can result in electron-positron pair production \citep{1967PhRv..155.1404G}, if the energy exceeds the threshold condition, given by
\begin{equation}
  \epsilon \epsilon_\gamma (1-\cos \theta_\mathrm{int}) \geq 2,
  \label{eqn:gaga_threshold}
\end{equation}
where the energies are normalized to the electron rest-mass energy, i.e. $\epsilon = h\nu/m_{\rm e}c^2$, and $\theta_\mathrm{int}$ is the interaction angle between the two photons. 

The \gaga\ optical depth is given by \citep{1967PhRv..155.1404G}
\begin{equation}
  \tau_{\gamma\gamma} = \int_0^l\,{\rm d}l \int_{4\pi}  {\rm d}\Omega \, \,(1-\mu) \int_{\frac{2}{\epsilon_\gamma (1-\mu)}}^\infty {\rm d} \epsilon \, n_{\rm ph}(\epsilon,\Omega) \sigma_{\gamma\gamma}(\epsilon,\epsilon_\gamma,\mu)
\label{eqn:tau_gamma}
\end{equation}
where $l$ is the distance over which the \gammaray\ photon travels, 
$\mu = \cos \theta_\mathrm{int}$, ${\rm d}\Omega = \sin{\theta}d\theta d\phi$
is the solid angle element in the spherical coordinate system centered at the gamma-ray photon with zenith determined by the direction from the star (see Appendix~\ref{a:num}), and $n_{\rm ph}(\epsilon,\Omega)$ is the number density per unit solid angle of the low-energy target photons. Here, $\sigma_{\gamma\gamma}$, is the \gaga\ cross-section \citep{1976tper.book.....J},
\begin{equation}
 \sigma_{\gamma\gamma}(\beta) = \frac{3}{16} \sigma_{\rm T} ( 1- \beta^2) \left[ (3-\beta^4) \ln \left(\frac{1+\beta}{1-\beta} \right) -2\beta (2-\beta^2)\right],
\end{equation}
where 
\begin{equation}
 \beta = \sqrt{1-\frac{2}{\epsilon \epsilon_\gamma (1-\mu)}},
\end{equation}
and $\sigma_{\rm T}$ is the Thomson cross-section. 

The radiation field, which provides the target photons for inverse Compton scattering and \gaga\ absorption, is produced by the massive companion star. In the case of a Be star companion, the photon field is produced by two components: the optical stellar radiation and the infrared radiation from the circumstellar disc. 
Calculations of the \gaga\ absorption in \psrb\ due to the circumstellar disc \citep{2017ApJ...837..175S} showed that the circumstellar disc component is considerably less important than the stellar radiation field. The energy densities of the two radiation fields are constrained by the observed luminosities of the star and of the circumstellar disc. Although the energy density of the circumstellar disc photons can be higher the effective interaction length is much smaller and hence the contribution to the overall \gaga\ absorption is insignificant.
In this study for the simplification of the method we only take into account stellar photons. We then approximate the photon density distribution by black-body radiation and assume that photons are emitted radially. Therefore,
\begin{equation}
n_{\rm ph} (\nu,\Omega) = \frac{1}{h \nu c} B_\nu(T_\ast) ,
\end{equation}
where $h$ is the Planck constant, $c$ is the speed of light, $T_\ast$ is the effective temperature of the star, and $B_\nu(T_\ast)$ is the Planck function, i.e. the spectral power emitted per unit area per unit solid angle. 

It is clear from Eq.~\ref{eqn:tau_gamma} that for a known energy and angular distribution of target photons the optical depth would depend exclusively on the energy and trajectory of a gamma-ray photon. The trajectory of a gamma-ray photon traveling along the line of sight in turn depends on the the geometry of the system and the location of the emitting region at a particular moment of time.
The location of the emitting region or region characterized by the most effective particle acceleration in gamma-ray binaries is unclear and is still under active debate. 
The emitting location will be different depending on if the compact object is a pulsar or a black hole. In the former case particles are believed to be accelerated at the termination shock between the pulsar and stellar winds as well as at other shocks generated due to the winds interaction \citep[see e.g.][]{2015A&A...581A..27D,2020A&A...633A.102H, 2021A&A...646A..91H}, while in the latter \add{case} particles can be accelerated anywhere along the jet \citep[see e.g.][]{2006ApJ...643.1081D,2009IJMPD..18..347B}. Moreover, it is not completely clear which part of the termination shock (or other shock) in the wind-driven scenario
would be the most favourable for particle acceleration and this can also change \is{with} orbital phase of the pulsar (see Section~\ref{sec:intro}). It should be noted that this can have a dramatic impact on the expected emission spectrum as well as the expected gamma-gamma absorption. In the wind-driven scenario the apex of the bow shock is argued to dominate the gamma-ray emission. In this case \gaga\ absorption could have a strong effect due to the proximity to the companion star. But if the gamma-ray emission is produced at the back (Coriolis) or secondary shocks the emission region could be too far from the star for any significant pair production to take place. Similarly, for microquasars, the effect of \gaga\ absorption would be very different depending on whether the emission is produced at the base of the jet or further along the jet.
%
%

In the following we eliminate this effect from our considerations by assuming a point-like scenario, i.e. that particles are accelerated, and hence the gamma-ray emission is produced, close to the compact object. We acknowledge that this assumption would be realistic only for the accretion powered systems where gamma-ray emission is generated at the base of the jet being launched close to the central engine. For the purposes of the proposed method this assumption is, however, also well justified for wind driven systems with the pulsar as a compact object. Depending on where exactly the dominant fraction of the gamma-ray emission is produced (apex or tail of the shock) the level of the \gaga\ absorption will change, but it will not significantly impact the orbital dependence of the optical depth and specifically the orbital phase of the maximum absorption, which is of prime interest for the proposed method, because the location of the emitting region is in (or close to) the orbital plane. This was demonstrated for the case of \psrb\ in \citet{2017ApJ...837..175S} where the optical depth was calculated both assuming the gamma-ray emission is produced at the location of the pulsar and at the location of the apex of the shock. In some systems the massive companion star features a circumstellar disc which would push the wind termination shock closer to the pulsar during disc crossings. In this case the apex of the shock would actually be closer to the location of the pulsar. 
The situation can be more complicated if the dominant site of the gamma-ray emission production is changing with the orbital phase. In this case the proposed method would not be applicable as the dependence of the optical depth on the orbital phase can be regulated by how the location of the emission region is changing. This scenario is, however, not well understood and beyond the scope of this paper. Finally, this method would also not be applicable for the microquasar scenario where the gamma-ray emission is produced along the jet at a considerable distance from the black hole. In this case the dependence of the \gaga\ absorption on the orbital phase can be very different and would depend on the orientation of the jet. 
%

\begin{figure*}
\centering  
  \includegraphics{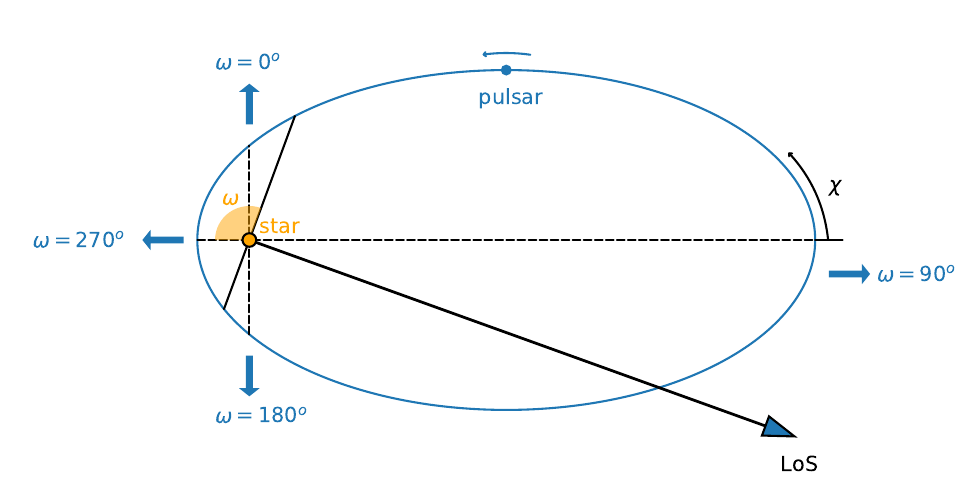}
  \caption{Schematic cartoon depicting the geometry of the binary system projected on the orbital plane. The LoS direction shows the line-of-sight towards the observer, $\chi$ is the angular orbital phase measured from apastron and $\omega$ is the longitude of periastron. For clarity we also show on the plot what the value of $\omega$ would be if the LoS direction is aligned with the blue arrows. }
  \label{fig:geom_cartoon}
\end{figure*}

Adopting the assumptions above, we end up with the optical depth being exclusively a function of the geometry of the system, i.e. the inclination angle $i$ (the angle between line-of-sight (LoS) and the normal to the orbital plane), the longitude of periastron $\omega$ (the angle between the ascending node and periastron), the eccentricity $e$, and the angular orbital phase $\chi$ which we choose to measure from apastron in the direction of motion of the compact object (true anomaly minus $\pi$; see Fig.~\ref{fig:geom_cartoon}). Further, following the hypothesis that the orbital phase of the characteristic dip in the TeV light curve of a binary is defined by the strongest absorption, we can probe the geometry of the system.  

\subsection{Analytic considerations}
\label{sec:analytic}

\is{Before performing full numerical calculations, we present the derivation of a simple analytic approximation of the \gaga\ optical depth} as a function of the orbital phase $\chi$ measured from apastron in the direction of the movement (see Fig.~\ref{fig:geom_cartoon}). For this we approximate the \gaga\ cross section by \citep{1985ApJ...294L..79Z}: 
\begin{equation}
    \sigma_{\gamma\gamma} = \frac{1}{3}\sigma_{\rm T}\epsilon \, \delta(\epsilon - \frac{2}{\epsilon_\gamma}).
\end{equation}
eliminating the dependence of the cross section on the energy of the soft photon and the interaction angle. Figure 3.12 in \citet{2012rjag.book.....B} shows the comparison of the pair production rate spectra using different approximations for the cross section. The delta-function approximation works very well for intermediate energies of resulting pairs but predicts artificially hard cutoffs compared to the full
expression of the cross section. Assuming the stellar black body radiation as the main target photon field and adopting a point-like approximation we can express the number density of the photon field as
\begin{equation}
    n_{\rm ph}\left(\epsilon, \theta\right) = m_{\rm e}c^2 \frac{8\pi}{h^3c^3} \left(\epsilon m_{\rm e}c^2\right)^2 \frac{1}{e^{\epsilon m_{\rm e}c^2/kT} - 1} \frac{R_\ast^2}{r(x)^2} \delta\left(\theta - 0\right)
\end{equation}
where $r$ is the distance from the gamma-ray photon to the star and $x$ is the distance the gamma-ray photon has moved along the line of sight. The delta-function reflects the assumption that the stellar photons are radiated radially from the star taking into account that the zenith is aligned with the direction from the star to the gamma-ray photon as described above.
Then, taking into account both delta-functions Eq. \ref{eqn:tau_gamma} can be rewritten as
%
\begin{equation}
    \label{eqn:taugg_approx}
    \tau_{\gamma\gamma}(\epsilon_\gamma,\chi) = \Phi(\epsilon_\gamma) \int_0^{2\pi}{\rm d}\phi \int_0^l {\rm d}x \, \frac{1-\mu(x,\chi)}{r(x,\chi)^2} = 2\pi\Phi(\epsilon_\gamma) f(\chi),
\end{equation}
where  
\begin{equation}
    \Phi(\epsilon_\gamma) = \frac{64\pi}{3 h^3 c^3} \sigma_{\rm T} \left( \frac{m_{\rm e} c^2}{\epsilon_\gamma} \right)^3 \frac{R_\ast^2}{e^{2 m_{\rm e}c^2/\epsilon_\gamma kT} - 1}
    \label{eqn:f_phi_int_form}
\end{equation}
\is{is only a function of the gamma-ray energy and is constant with the orbital phase. The }
function 
\begin{equation}
f(\chi) = \int_0^l {\rm d}x \, \frac{1-\mu(x,\chi)}{r(x,\chi)^2}
\label{eqn:f_chi_integral}
\end{equation}
reflects the geometry of the system and determines the variability of the optical depth with the orbital phase.
The terms $\mu(x,\chi)$ and $r(x,\chi)$ can be expressed as functions of the distance between the star and gamma-ray emitting region $d$ and the initial interaction angle, $\mu_0 \equiv \mu(x = 0)$, as 
\begin{equation}
    \label{eqn:mu}
    \mu(x,\chi) = \sqrt{\left(1 - \frac{d(\chi)^2(1 - \mu_0(\chi)^2)}{r(x,\chi)^2}\right)}, 
\end{equation}
\add{and}
\begin{equation}
    \label{eqn:r}
        r(x, \chi)^2 = d(\chi)^2 + x^2 + 2xd(\chi)\mu_0(\chi).
\end{equation}
As long as the observer is far enough away from the binary system, the integration in 
Eq.~\ref{eqn:f_chi_integral} 
can be performed up to infinity. Then substituting Eqs.~\ref{eqn:mu} and \ref{eqn:r} $f(\chi)$ in Eq.~\ref{eqn:f_chi_integral} and performing the integration we find that $f(\chi)$ can be expressed analytically as (see Appendix~\ref{a:fphi} for details)
\begin{equation}
    \label{eqn:f_phi}
    f(\chi) = \frac{1}{d(\chi)}\left[ \frac{1}{\sqrt{1 - \mu_0(\chi)^2}} \left(\frac{\pi}{2} - \arctan{\frac{\mu_0(\chi)}{\sqrt{1 - \mu_0(\chi)^2}}}\right)  - 1  \right].
\end{equation}
The maximum of this function determines the orbital phase where the maximum gamma-gamma absorption occurs and hence, according to our hypothesis, the orbital phase of the dip in the TeV light curve.

Now, applying a point-like source approximation, i.e. the emission region is located at the position of the compact object, we can express $d(\chi)$ and $\mu_0(\chi)$ in terms of the orbital parameters of the system: 
\begin{equation}
    \mu_0(\chi) = \sin{i}\sin{(\chi+\omega)} ,
\end{equation}
\begin{equation}
    d(\chi) = \frac{a(1-e^2)}{1-e\cos{\chi}}.
\end{equation}
Following these simple considerations it can be seen that the maximum of the function $f(\chi, i, \omega, e)$ in Eq.~\ref{eqn:f_phi} is determined by the geometry of the binary system, namely its inclination angle, longitude of periastron, and eccentricity. Below we will demonstrate that this approximation is in fairly good agreement with the exact numerical calculations for a wide parameter space and hence can be used to place rough constraints on the geometry of the binary system.

\subsection{Comparison of analytic and numeric solutions}

\begin{figure*}
\centering
\begin{subfigure}
         \centering
         \includegraphics[width=0.3\textwidth]{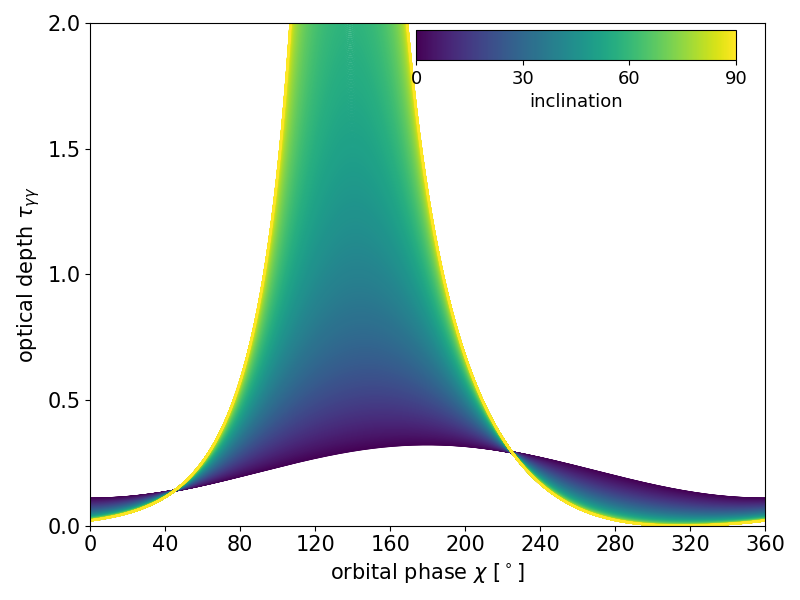}
\end{subfigure}
\begin{subfigure}
         \centering
         \includegraphics[width=0.3\textwidth]{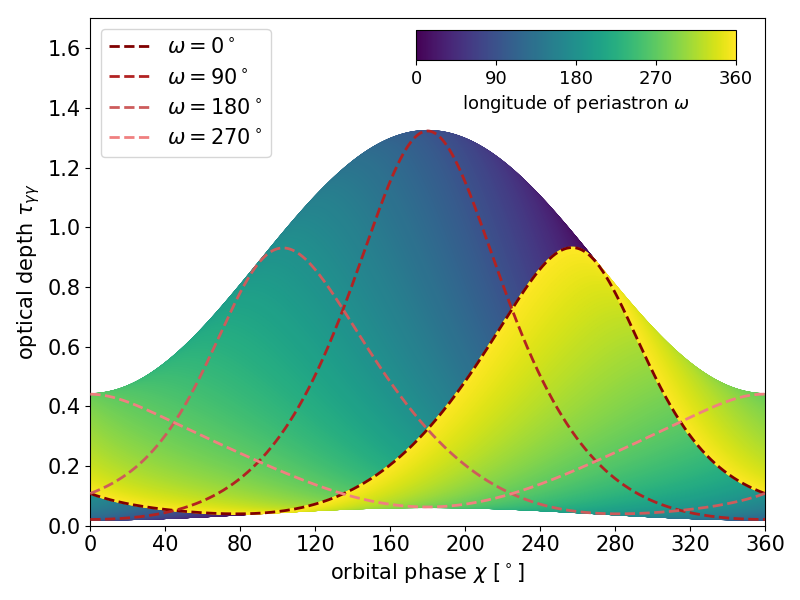}
\end{subfigure}
\begin{subfigure}
         \centering
         \includegraphics[width=0.3\textwidth]{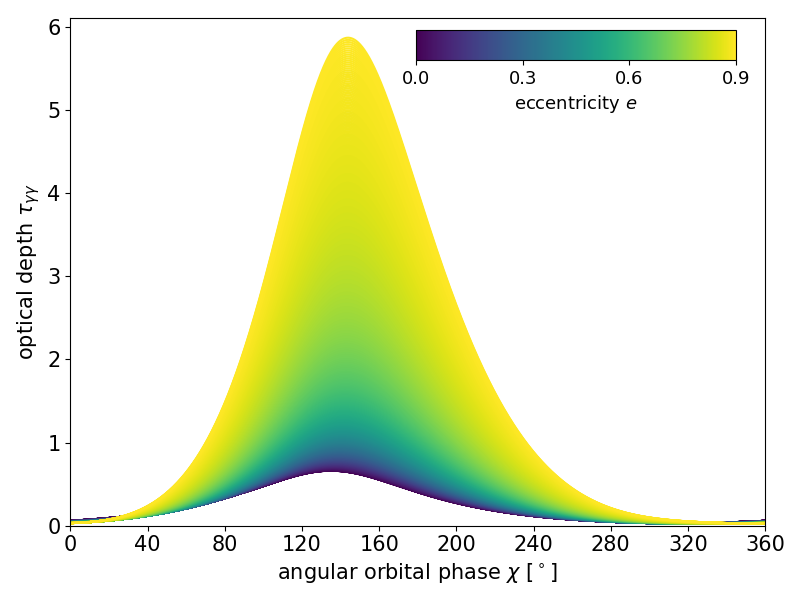}
\end{subfigure}
\begin{subfigure}
         \centering
         \includegraphics[width=0.3\textwidth]{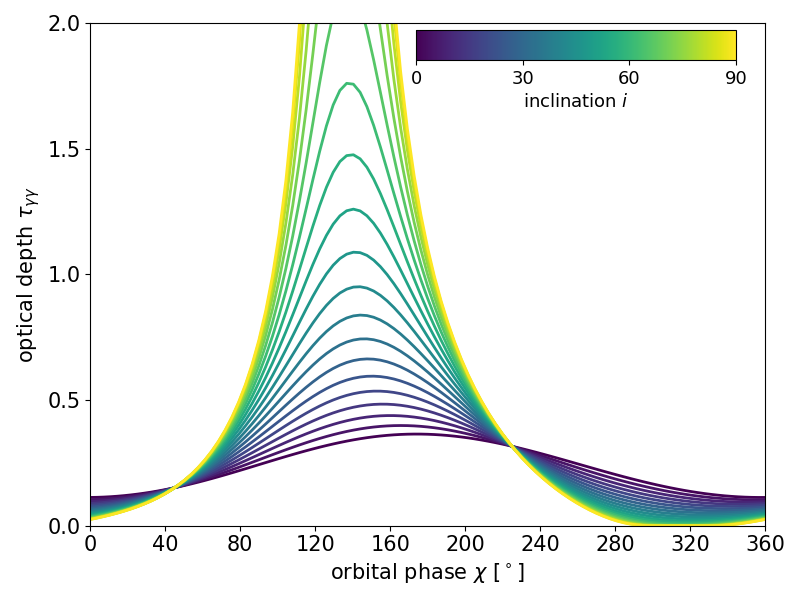}
\end{subfigure}
\begin{subfigure}
         \centering
         \includegraphics[width=0.3\textwidth]{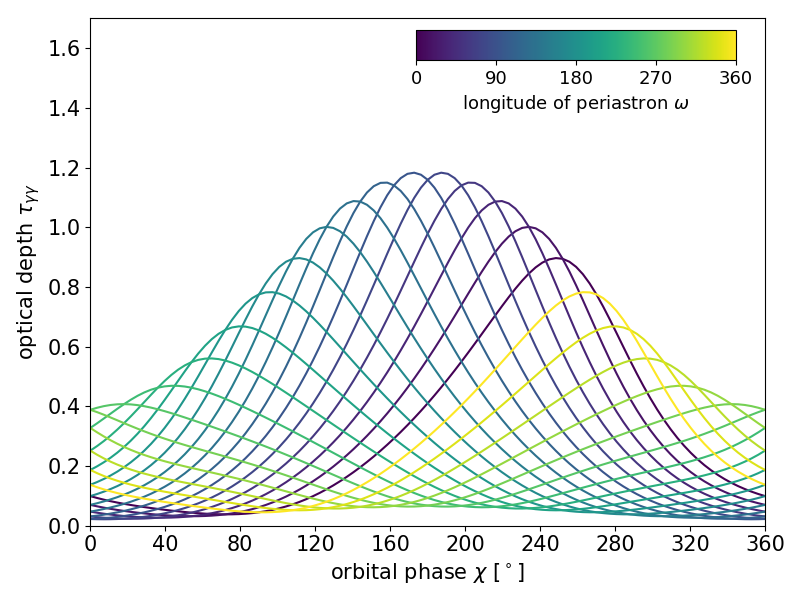}
\end{subfigure}
\begin{subfigure}
         \centering
         \includegraphics[width=0.3\textwidth]{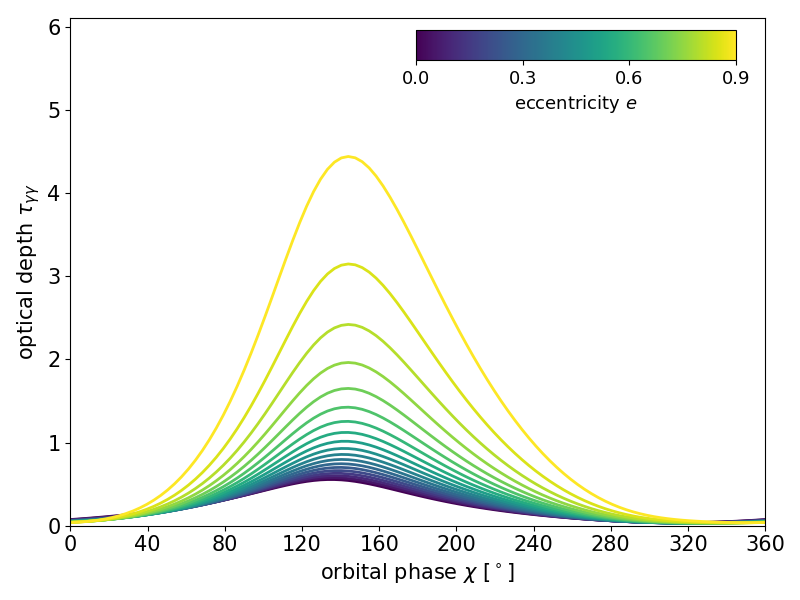}

\end{subfigure}

\begin{subfigure}
         \centering
         \includegraphics[width=0.3\textwidth]{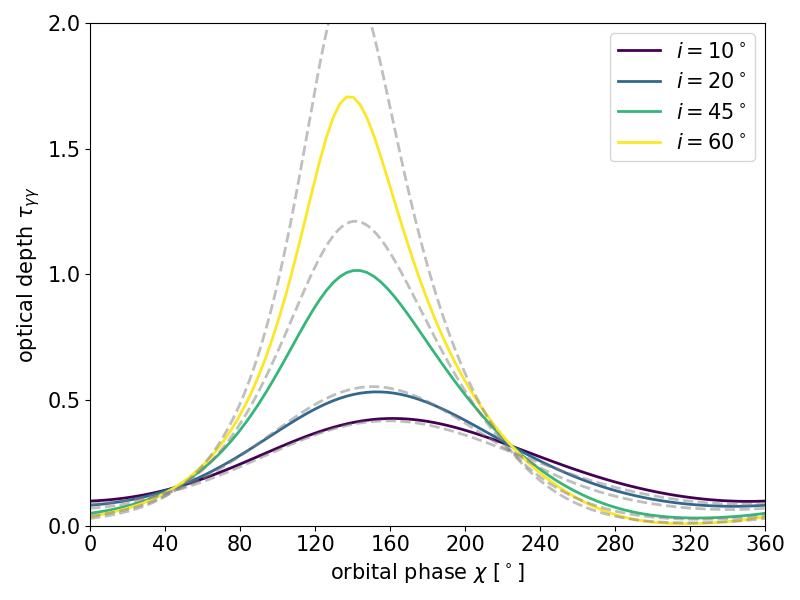}
\end{subfigure}
\begin{subfigure}
         \centering
         \includegraphics[width=0.3\textwidth]{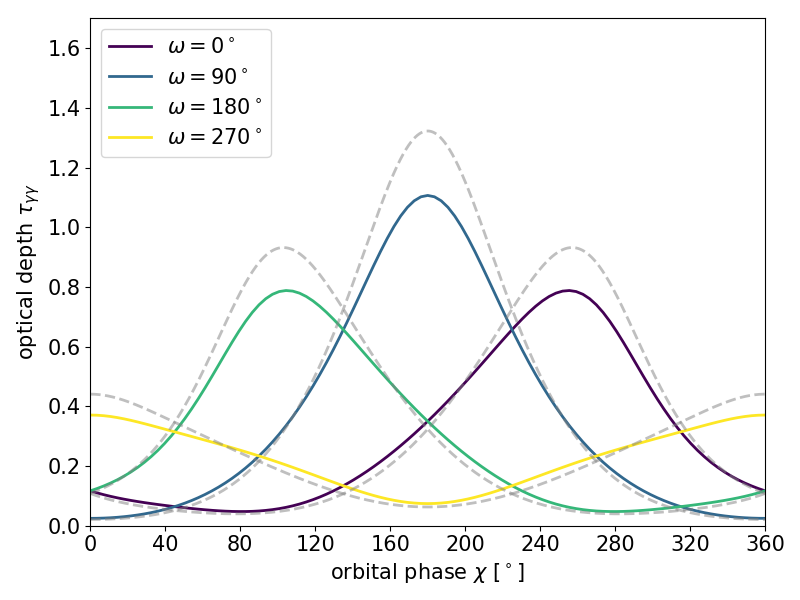}
\end{subfigure}
\begin{subfigure}
         \centering
         \includegraphics[width=0.3\textwidth]{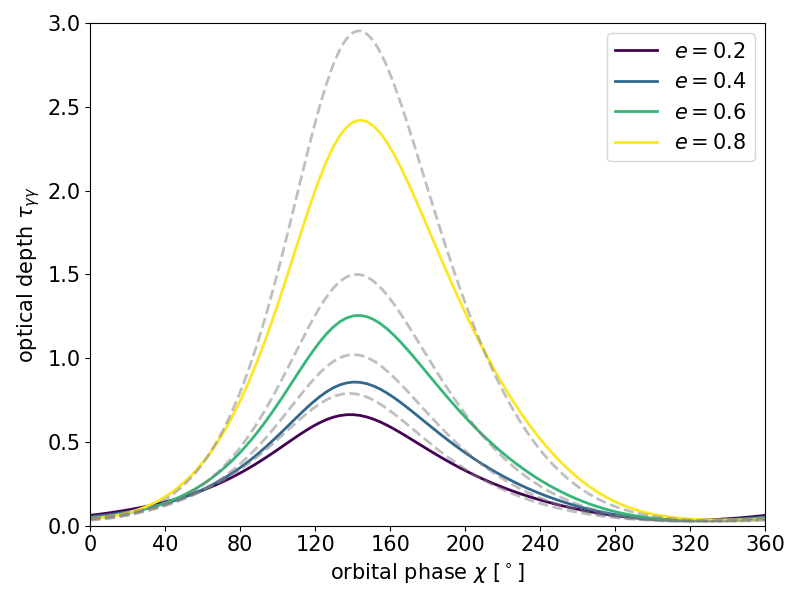}

\end{subfigure}

\caption{Optical depth $\tau_{\gamma\gamma}$ as a function of the orbital phase calculated analytically (top panels) and numerically (middle panels) for varying values of inclination (left panels) longitude of periastron (middle panels) and eccentricity (right panels). A direct comparison of the optical depth calculated analytically (dashed lines) and numerically (solid lines) for different values of corresponding parameters is shown in the bottom panels.}
\label{fig:parameter_scan}
\end{figure*}

Numeric calculation of the optical depth were done following the approach presented in \citep{2017ApJ...837..175S} which for the stellar radiation component essentially follows \citet{2006A&A...451....9D}. The method is summarized in the Appendix~\ref{a:num}. To explore the deviation of the analytic approximation from the exact numeric simulation we consider a Test Binary with parameters listed in the Table~\ref{tab:binaries}. The optical depth is calculated both analytically and numerically varying separately the inclination, the longitude of periastron and the eccentricity. Other parameters are kept fixed. In Figure \ref{fig:parameter_scan} we show the \gaga\ optical depth $\tau_{\gamma\gamma}$ as a function of the orbital phase $\chi$ with color coding indicating the range of varied parameters: left panels - varied inclination with the longitude of periastron fixed at $135^\circ$ and eccentricity fixed at $0.5$; middle panels - varied longitude of periastron with the inclination fixed at $45^\circ$ and the eccentricity fixed at $0.5$; and right panels - varied eccentricity with the inclination fixed at $45^\circ$ and the longitude of periastron fixed at $135^\circ$. The upper panels show the analytical approximation discussed above, the middle panels show the exact numeric solutions of Eq.~\ref{eqn:tau_gamma}, and the lower panels \add{show direct comparisons for some selected parameters. }

In general, \is{there is} a rather good agreement between the analytic and numeric solutions. Both solutions show the same evolution of the shape of the optical depth curve 
with the change of the orbital parameters. The analytic approximation slightly overestimates the value of the optical depth which becomes more significant at large inclination angles. This, however, does not play a major role for the purpose of this paper. 

\begin{figure}
    \centering
    \includegraphics[width=\hsize]{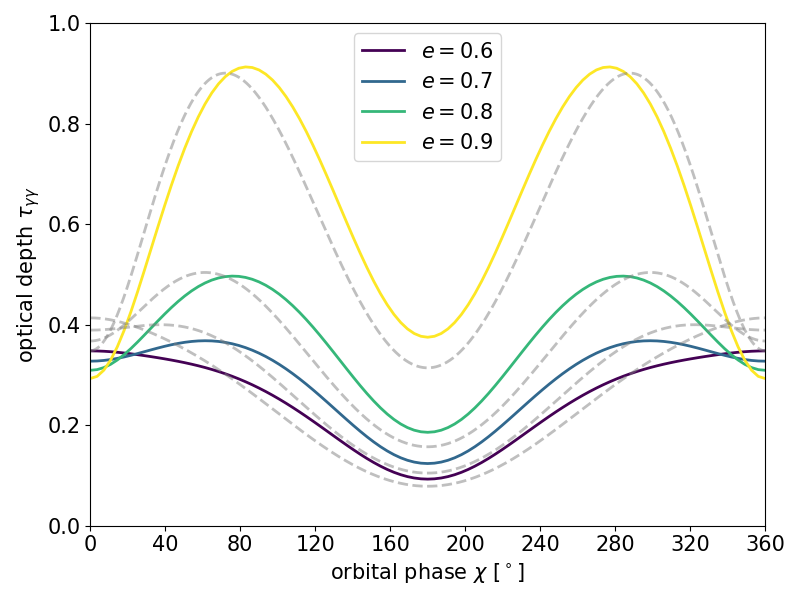}
    \caption{Same as the right bottom plot in Fig.~\ref{fig:parameter_scan} but for $\omega = 270^\circ$. }
    \label{fig:w270}
\end{figure}

A careful parameter scan showed that the discrepancy between the analytic approximation and numeric calculations in determination of the orbital phase of \is{maximum absorption} becomes more significant for a specific case of $\omega \sim 270^\circ$, i.e. the periastron coincides with the inferior conjunction. \bvs{This is particularly true for high eccentricities} where the optical depth curve exhibits two maxima for $\omega \sim 270^\circ$ (Fig.~\ref{fig:w270}).

\begin{table*}[]
\centering
\caption{Physical parameters of examined systems.}
\label{tab:binaries} 
\setlength\tabcolsep{0pt} 
    \begin{tabular*}{\textwidth}{@{\extracolsep{\fill}} *{5}{c}}
    \hline
    \hline
    &Test Binary& PSR B1259-63\ding{75} & LS 5039\ding{70} & PSR J2032$+$4127\ding{61}\\
    \hline
    $i$ [$^\circ$]& 45 &22.2& 60 & $>27.7$\\
    $\omega$ [$^\circ$] \ding{64} & 135 &  138.7 & 45.8 (56.0) &  $21-52$\\
    $e$ & 0.5&0.87 & 0.35 (0.337) &$0.936-0.989$\\
    $P$ [days]& 100 & $1,236.72$ & 3.906&$16,000-17,670$\\
    \hline
    $T_\ast$ [K]& 30,000& 33,000& 39,000 & 20,000 \\
    $M_\ast$ [M$_\odot$]& 30 &31  & 22.9 & 15\\
    $R_\ast$ [R$_\odot$]& 10 & 9.2 & 9.3&  10\\
    \hline
    \end{tabular*}
\vspace{5pt}
\raggedright
\\
    \footnotesize{\ding{64} Longitude of periasrton of the pulsar orbit}\\  
    \footnotesize{\ding{75} Orbital and stellar parameters are adopted from \citet{2011ApJ...732L..11N} and references therein}\\
    \footnotesize{\ding{70} Orbital and stellar parameters are adopted from \citet{2005MNRAS.364..899C}; alternative orbital parameters (in brackets) are from  \citet{2009ApJ...698..514A}}\\
    \footnotesize{\ding{61} \citet{2017MNRAS.464.1211H} for orbital parameters and \citet{2015MNRAS.451..581L} for stellar parameters}\\
\end{table*}

\begin{figure}
    \centering
    \begin{subfigure}
        \centering
        \includegraphics[width=\hsize]{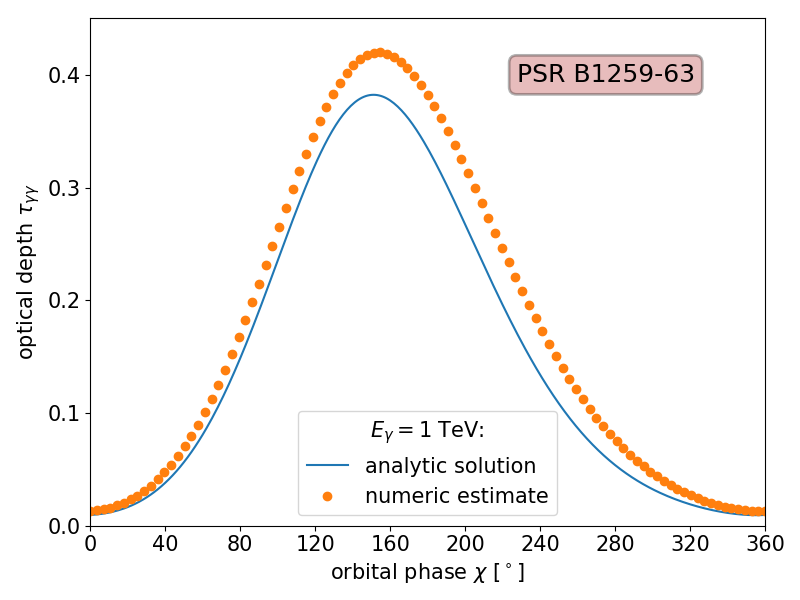}             
    \end{subfigure}

    \begin{subfigure}
        \centering
        \includegraphics[width=\hsize]{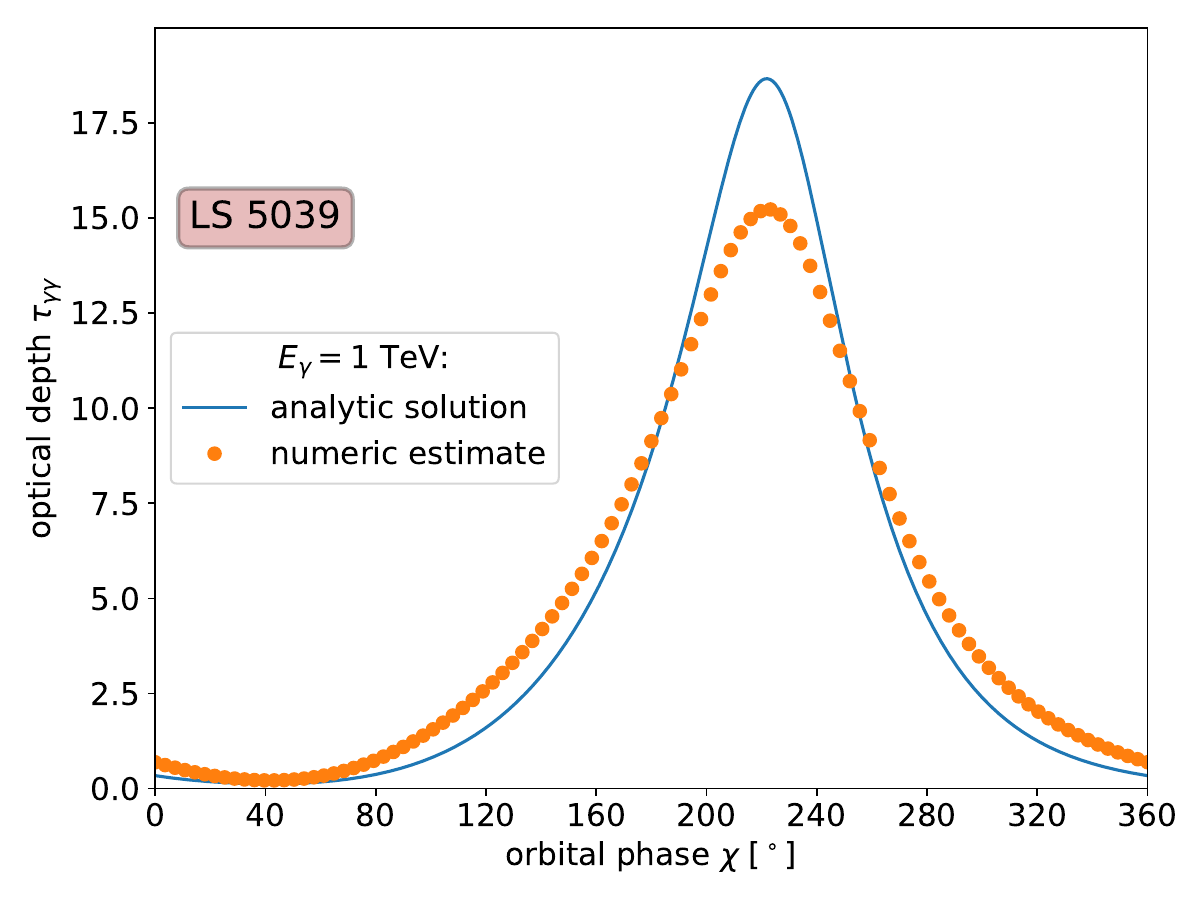}         
    \end{subfigure}

    \caption{The optical depth for gamma-gamma absorption of the 1 TeV gamma-ray photon as a function of the orbital phase from the binary system PSR~B$1259-63$ (top panel) and LS~5039 
    \citep[bottom panel; orbital solution by][]{2005MNRAS.364..899C}. Filled orange circles represent the exact numeric estimate of Eq.~\ref{eqn:tau_gamma}, while the solid blue line shows the analytic approximation derived in Section~\ref{sec:analytic}.}
    \label{fig:analytic_psrb1259}
\end{figure}

Figure~\ref{fig:analytic_psrb1259} shows the application of the optical depth calculation to two well studied binary systems, PSR~B$1259-63$ (top panel) and LS~5039 (bottom panel), with substantially different orbital parameters (Table~\ref{tab:binaries}). While PSR~B$1259-63$ is a very eccentric and long period binary, LS~5039 is a compact binary with a low eccentricity. In both cases the analytic approximation reproduces the shape of the numerically calculated curve relatively well \is{with a good match of the orbital phase of the peak optical depth}.

\subsection{The orbital phase of the maximum absorption}
\label{sec:maxphase}



To estimate the orbital phase where the gamma-gamma absorption is the strongest and hence where the dip in the observed light curve is expected we vary the geometrical parameters of the orbit, namely inclination between 0 and 90 degrees (20 bins) and longitude of periastron between 0 and 360 degrees (20 bins), and for each combination of $i$ and $\omega$ we calculate the orbital phase with the highest 
optical depth, i.e. maximum absorption, $\phi_\mathrm{max}$. 
For the results presented below we use exact numeric simulations
\bvs{but the analytical approximation is compatible with these results.}
The dependence of the maximum absorption phase on the orbital parameters is further discussed and a detailed comparison of the numeric and analytic approach is presented in Appendix \ref{a:anVSnum}. For numeric calculation we split the orbit into 200 bins equidistant in orbital phase setting a systematic error of of $1.8^\circ$ on the estimate of $\phi_\mathrm{max}$.

\section{Application to PSR J2032+4127/MT91 213}
\label{sec:tev2032}

PSR J$2032+4127$/MT91 213 is only the second known gamma-ray binary for which it is well established that the compact \add{object} is a pulsar \citep{2018ApJ...867L..19A}. The system consists of the pulsar PSR J$2032+4127$, which was first detected at GeV energies by Fermi-LAT \citep{2009Sci...325..840A} and later confirmed in radio observations \citep{2009ApJ...705....1C}, and a $\sim15M_\odot$ Be star MT$91 213$ \citep{2015MNRAS.451..581L}. The periastron passage in 2017 presented us with a unique opportunity to detect this binary system at VHE which otherwise would be impossible given its extremely eccentric orbit ($e>0.93$) and long \add{orbital} period of about $45-50$ years \citep{2017MNRAS.464.1211H}. The binary benefited from an extensive coverage across the electromagnetic regime during its periastron passage \citep{2018ApJ...867L..19A, 2019MNRAS.485.1864C, 2019ApJ...880..147N, 2019MNRAS.485.1864C},
which undoubtedly provided a better understanding of the system and allowed some constraints to be set on the physical parameters. However, PSR J$2032+4127$/MT91 213 still suffers from a poorly constrained orbit. The only available orbital solution provided by \citet{2017MNRAS.464.1211H} contains a lot of uncertainties refining the orbital period in the range $16000-17670$ days, eccentricity in the range $0.936-0.989$, longitude of periastron in the range of $21^\circ-52^\circ$ and the inclination angle above $27.7^\circ$ (taking into account the uncertainty on the mass function and on the mass of the Be star).

VHE observations of the PSR J$2032+4127$/MT91 213 binary system around the 2017 periastron passage were conducted by both VERITAS and MAGIC \citep{2018ApJ...867L..19A} \bvs{and revealed} a firm detection of the variable emission associated with the binary. The observed TeV light curve shows a rather typical behaviour with a two-bump structure and a dip shortly after periastron. 
It is interesting that the shape of the X-ray light curve is very different from the TeV light curve. It exhibits a much broader dip with the dimming starting already $\sim40$~days before periastron and the minimum flux period spanning from shortly before to $5-10$ days after periastron \citep{2018ApJ...867L..19A}. Although the TeV dip is coincident with the minimal X-ray flux, so is the first TeV bump which occurs roughly at periastron where the X-ray flux is already at its minimum. 
These differences indicate that the decrease of X-ray and gamma-ray flux could be attributed to different processes which agrees with the gamma-gamma absorption scenario for the attenuation of the TeV flux.

\begin{figure}[t]
    \centering
    \includegraphics[width=\hsize]{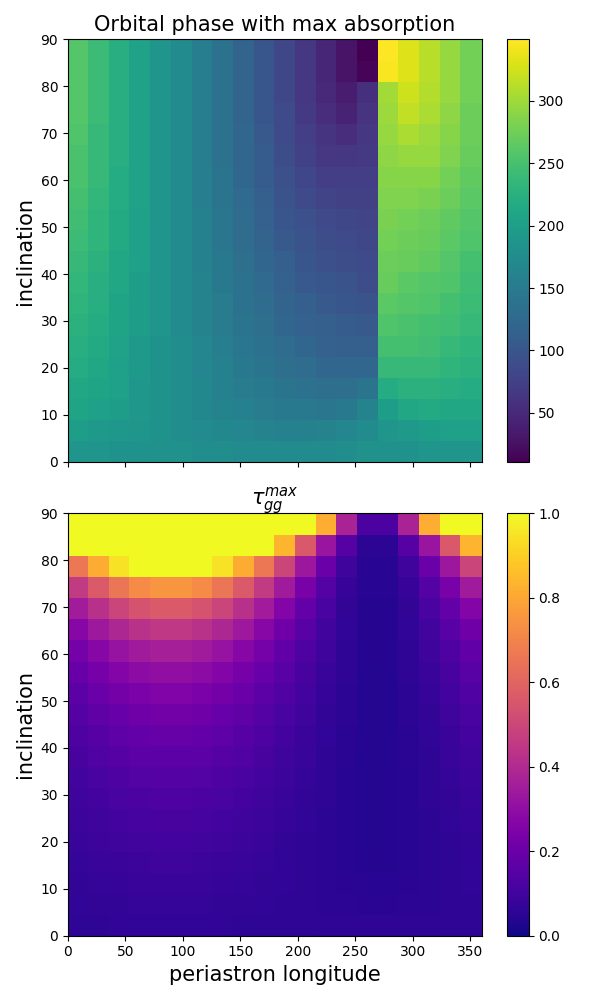}
    \caption{The color maps shows the orbital phase of the pulsar at which the emitted gamma-ray photon would encounter the highest gamma-gamma absorption along the line of sight as a function of assumed inclination and longitude of periastron \add{(top panel) as well as the value of the maximum optical depth (bottom panel)}.}
    \label{fig:param_search}
\end{figure}

To probe the orbital parameters of PSR J$2032+4127$/MT91 213 we follow the procedure described in Section~\ref{sec:maxphase}. In Fig.~\ref{fig:param_search} we show numerically calculated maps depicting the orbital phase where the maximum absorption occurs (top panel) and the value of the optical depth at that orbital phase (bottom) for the full parameter space of $i$ and $\omega$. The spread of potential maximum absorption locations covers almost the whole orbit depending on the inclination and the longitude of periastron. However, only for high inclincation angles is the optical depth high enough to cause significant absorption. It should be noted here that the assumption of the emitting region being located \is{at the pulsar position} might somewhat underestimate the optical depth. For these simulations we used an eccentricity of $e=0.961$ and a period of $P=17000$ days, which is the average `model 2' presented in \citet{2017MNRAS.464.1211H}. According to simulations presented in Fig.~\ref{fig:parameter_scan} small variations in the eccentricity do not strongly impact the estimate of the orbital phase with the maximum absorption. They can, however, significantly change the estimate of the optical depth itself, \is{resulting in a stronger absorption at $\phi_\mathrm{max}$ for higher eccentricity.}
\is{The estimate of $\phi_\mathrm{max}$ is not very sensitive to the orbital period as long as it is long enough and the size of the star is negligible (see Appendix \ref{a:anVSnum}).}
\add{For other parameters fixed, a short orbital period would result in stronger} absorption as the distance between the pulsar and the star is smaller.

Both MAGIC and VERITAS observations \citep{2018ApJ...867L..19A} indicate that the minimum in the light curve occurs at about 7 days after the periastron passage. In our diagnostic we conservatively consider a larger time interval of 8 days spanning from 3 days to 11 days after periastron. Conversion of days to the orbital phase is naturally very sensitive to the eccentricity 
and the \bvs{orbital period, estimates of which suffer from quite large uncertainties, for PSR J$2032+4127$/MT91 213.}

\begin{figure}
    \centering
    \begin{subfigure}
        \centering
        \includegraphics[width=\linewidth]{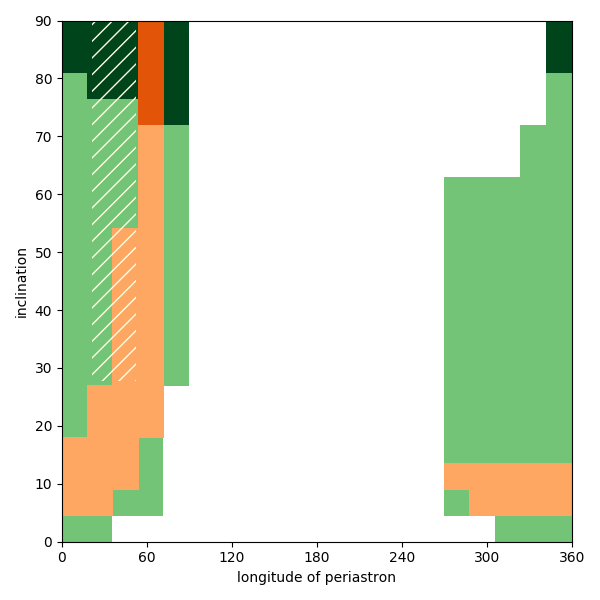}
    \end{subfigure}
    \begin{subfigure}
        \centering
        \includegraphics[width=\linewidth]{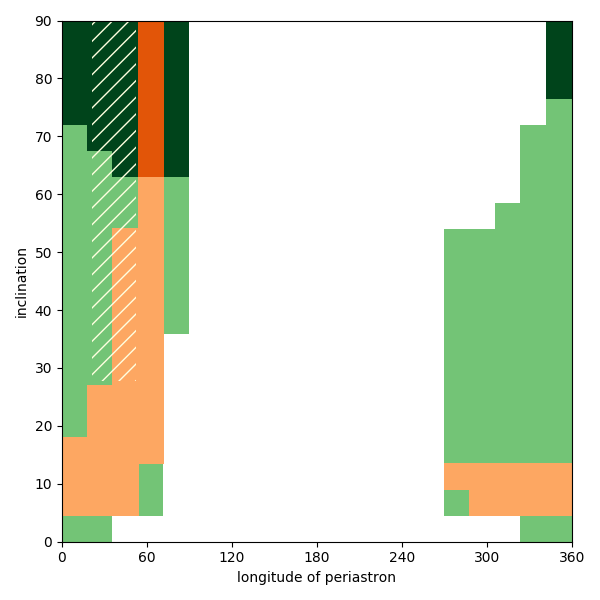}
    \end{subfigure}
    \caption{Allowed combinations of values of $i$ and $\omega$ as derived from the position of the minimum in the TeV light curve of PSR J$2032+4127$/MT91 213 that takes place at $7\pm4$ days after the periastron using numeric calculations (top panel) and analytic approximation (bottom panel). The green colors take into account the whole range of uncertainties for the estimates of the eccentricity and the period while orange colors assume the 'model 2' values of $e=0.961$ and $P=17000$ days from \citep{2017MNRAS.464.1211H}. Dark orange and dark green correspond to the region where the optical depth is high enough to attenuate the intrinsic flux by at least a factor of 2. Light yellow hatching represents the constraints on $i$ and $\omega$ set by optical observations \cite{2017MNRAS.464.1211H}}.
    \label{fig:allowed_values}
\end{figure}

Taking into account the whole range of \add{values of $e$ and $P$} allowed by optical observations 
\is{\citep[see Table \ref{tab:binaries};][]{2017MNRAS.464.1211H}} 
we end up with quite a large range of orbital phases, $186^\circ - 290^\circ$, which could correspond to the minimum in the light curve. Nevertheless, even this rather wide range of orbital phases is still quite constraining in the determination of the inclination angle and the longitude of periastron. In Fig.~\ref{fig:allowed_values} the regions marked with green colors represent the parameter space allowed if the dip in the TeV emission is due to gamma-gamma absorption, i.e. those that fall into the $186^\circ - 290^\circ$ range, as implied by numeric calculations (top panel) and analytic approximation (bottom panel). One can see that it covers roughly half of the whole parameter space. However, taking into account that the optical depth needs to be high enough to provide a sufficient level of absorption, one can further strongly reduce the parameter space. The dark green color corresponds to the parameter space where at least half to the intrinsic radiation is absorbed. It is clear that taking into account the level of absorption strongly constrains the inclination, suggesting a highly inclined orbit. Naturally a better estimate of the orbital period and eccentricity further constrains the allowed parameter space. The orange region shows the parameters allowed by 'model 2' of \citep{2017MNRAS.464.1211H} that adopts average values of $P=17000$~days and $e=0.961$. This allows us to much better constrain the longitude of periastron. Again, the dark orange region reflects significant levels of absorption.

The dashed regions in Fig.~\ref{fig:allowed_values} illustrate the constraints on $i$ and $\omega$ as set by optical observations \citep{2017MNRAS.464.1211H}. Their comparison to the constraints obtained in this work can be summarized in a few main points:
\begin{itemize}
    \item a clear overlap of the regions with allowed values strongly supports the general idea that the location of the minimum in the TeV light curve is determined by the gamma-gamma absorption and reflects the geometry of the system
    


    \item a precise measurement of the orbital phase of the dip in the TeV light curve, that also requires a good knowledge of the orbital period and eccentricity,
    could potentially not only strongly constrain the $i-\omega$ parameter space but also yield quite precise estimates of at least one of these two parameters, namely the longitude of periastron.
    \item the estimate of the optical depth which is a measure of the level of absorption might strongly constrain the inclination of the system under condition of good understanding of where the emitting region is located.
    \item vice versa, in the case of reliable estimates of the orbital geometry from other considerations like optical observations or radio/gamma-ray pulsar timing measurements, the location of the dip in the TeV light curve can indirectly indicate the location of the emitting region and hence provide a valuable insight into understanding of the particle acceleration in these systems.
\end{itemize}
The latter is certainly the case for the \psrb\ binary and will be further investigated in our future works.

Comparison of the top (numeric calculations) and bottom (analytic approximation) panels in Fig.~\ref{fig:allowed_values} does not show a big difference between the two approaches implying 
that the analytic approximation could be a useful tool that can be used to constrain the geometry of the orbit without spending a lot of CPU hours.

\section{Summary}

In this study we examine the impact of gamma-gamma absorption on the periodic very-high-energy radiation from gamma-ray binaries and study how this impact depends on the orbital parameters of a particular system. We suggest that under the assumption that a dip-like feature in the TeV light curve (characterized by a fast decrease flux superposed on a general increase of flux) can be attributed to the highest gamma-gamma absorption one can infer constraints on orbital parameters of the system from the orbital location of this dip. We propose a method to diagnose TeV light curves and use them as a probe for orbital parameters. We also derive an analytic approximated solution for the optical depth that agrees reasonably well with the exact numeric solution and provides an efficient tool for quick diagnostics. Application to PSR J$2032+4127$/MT91 213 results in constraints on the inclination and longitude of periastron that are in a good agreement with constraints obtained from optical observations, implying that indeed this method could be used for characterization of the orbit. Moreover, we argue that precise time-resolved flux measurements as well as accounting for the level of absorption would further constrain the orbital parameters. Therefore, the method of gamma-gamma absorption offers another completely independent way of probing orbital parameters that could be complementary to classic optical \add{radial velocity} measurements. On the other hand, well determined orbital geometry combined with the TeV light curve could point to the location of the emitting region and shed light on acceleration and emission processes in these systems.

\begin{acknowledgments}
Iurii Sushch acknowledges support by the National Research Foundation of South Africa (Grant Number 132276). Brian van Soelen acknowledges support by the National Research Foundation of South Africa (Grant Number 119430). Some of the numerical calculations presented were performed using the University of the Free State High Performance Computing Unit.  
\end{acknowledgments}

%






\appendix
\section{Derivation of function $f(\phi)$}
\label{a:fphi}

Substituting Eqs.~\ref{eqn:mu} and \ref{eqn:r} $f(\phi)$ in Eq.~\ref{eqn:taugg_approx} and changing the upper limit of integration to infinity we get 
\begin{equation}
    \label{eqn:f_phi_int}
    f = \int_0^\infty \frac{1-\frac{x+d\mu_0}{\sqrt{d^2 + x^2 + 2xd\mu_0}}}{d^2 + x^2 + 2xd\mu_0}{\rm d}x.
\end{equation}

We further substitute $y = x + d\mu_0$ and split it into two integrals

\begin{align}
    f &= \int_{d\mu_0}^\infty \frac{{\rm d}y}{d^2(1 - \mu_0^2) + y^2} - \int_{d\mu_0}^\infty  \frac{y{\rm d}y}{(d^2(1 - \mu_0^2) + y^2)^{3/2}} \\
    &= I_1-I_2 \nonumber
\end{align}

For the first integral $I_1$,

\begin{align}
    \label{integral_one}    
    I_1 &=  \int_{d\mu_0}^\infty \frac{{\rm d}y}{d^2(1 - \mu_0^2) + y^2}  \nonumber \\
    &= \left. \frac{1}{d\sqrt{1 - \mu_0^2}} \arctan{\frac{y}{d\sqrt{1 - \mu_0^2}}}\right\rvert_{d\mu_0}^\infty  \\
    &=\frac{1}{d\sqrt{1 - \mu_0^2}} \left(\frac{\pi}{2} -  \arctan{\frac{\mu_0}{\sqrt{1 - \mu_0^2}}} \right). \nonumber
\end{align}

For the second integral $I_2$, conducting another change of variables, $t = y^2 + d^2(1 - \mu_0^2)$, we have

\begin{align}
    \label{integral_two}
    I_2 &= \frac{1}{2} \int_{d^2}^\infty t^{-3/2} {\rm d}t  = \left. - t^{-1/2}\right\rvert_{d^2}^\infty = \frac{1}{d}  
\end{align}

Combining \ref{integral_one} and \ref{integral_two} we get Eq.~\ref{eqn:f_phi}.

\section{Numeric simulations}
\label{a:num}

The integration over a solid angle in Eq.~\ref{eqn:tau_gamma} is performed in the spherical coordinate system centered at the location of the emitting region which is assumed to overlap with the location of the compact object. The zenith is determined by the direction from the center of the star to the emitting region. The integration in zenith angle can be then substituted by the integration in \revc{$\eta=\cos{\theta}$} and limited to \revc{$\eta\in \left[\cos{\theta_\ast}; 1 \right]$}, with $\theta_\ast$ being the angular radius of the star as viewed from the gamma-ray location, i.e. $\cos{\theta_\ast} = \frac{d}{\sqrt{d^2+R_\ast^2}}$, where $d$ is the distance between the gamma-ray location and the center of the star and $R_\ast$ is the radius of the star.
Additionally, we set the azimuth angle to be measured from the projected direction towards the observer and integrate over \revc{$\phi$} from 0 to $\pi$ using the symmetry. The solid angle integral in Eq.~\ref{eqn:tau_gamma} can be then rewritten as
\begin{equation}
\revc{\int_{4\pi}  {\rm d}\Omega = \int_0^{2\pi}d\phi\int_0^{\theta_\ast}\sin{\theta}d\theta = 2 \int_0^{\pi}d\phi\int_{\eta_\ast}^1 d\eta}
\end{equation}
\rev{where $\eta_\ast = \cos{\theta_\ast}$}. The integral along the line-of-sight is calculated in cylindrical coordinates centered on the center of the star and tied to the orbital plane from the location of the emitting region up to 1000 stellar radii.  

\begin{figure}
    \centering
    \begin{subfigure}
        \centering
        \includegraphics[width=0.48\hsize]{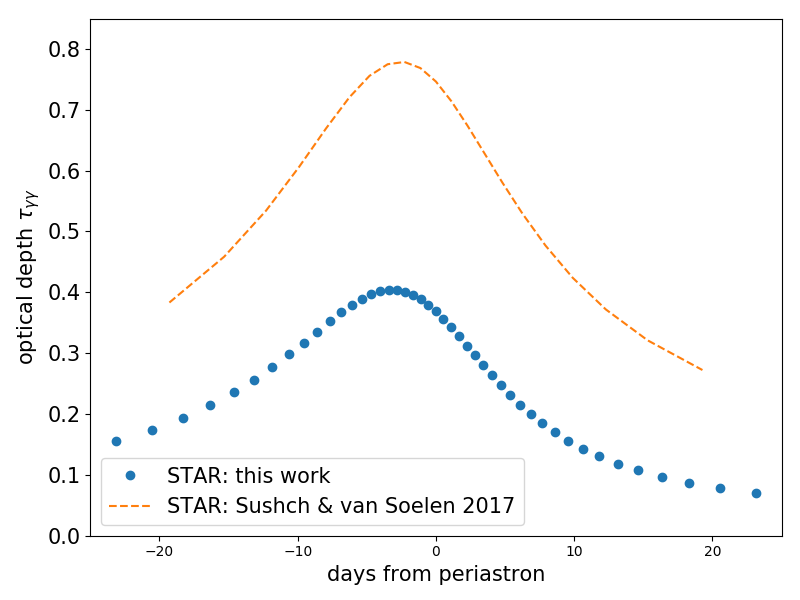}             
    \end{subfigure}
    \begin{subfigure}
        \centering
        \includegraphics[width=0.48\hsize]{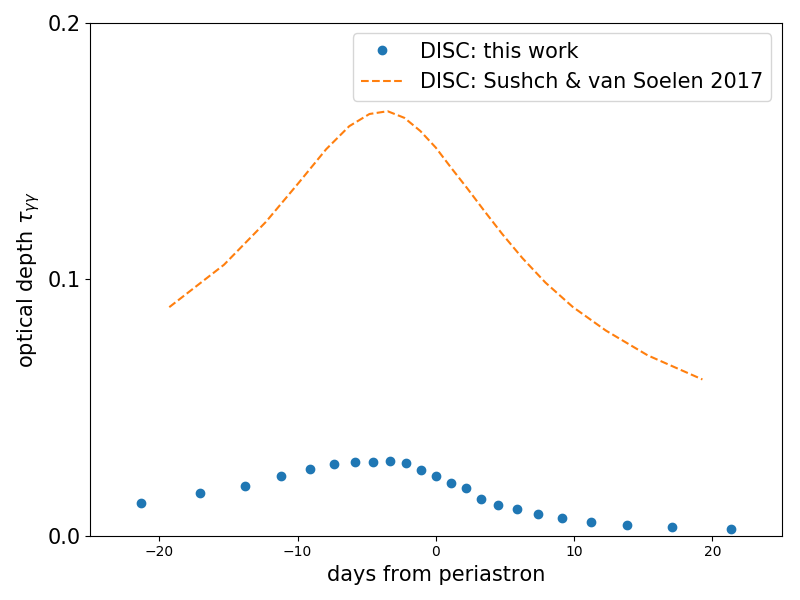}         
    \end{subfigure}
    \caption{The optical depth for gamma-gamma absorption of the 1 TeV gamma-ray photon by the stellar radiation field (left panel) and circumstellar disc (right panel) as a function of \add{days from periastron for} the binary system PSR~B$1259-63$ as calculated in \citet{2017ApJ...837..175S} with a geometric mistake (orange dashed line) and corrected in this work (blue filled circles). Note that \add{the calculation of optical depth due the stellar photons accounts for the} absorption of the stellar radiation by the circumstellar disc.}
    \label{fig:old_new}
\end{figure}

Note, that in \citet{2017ApJ...837..175S} the spherical coordinate system in which the integration over the solid angle is performed is defined differently due to accounting for the additional radiation field supplied by the circumstellar disk. It is centered at the emitting region but with zenith aligned with the normal to the disk. In calculations performed in \citet{2017ApJ...837..175S} we mistakenly flipped the direction of the soft photon to the opposite direction, i.e. the photon radiated from center of the star was calculated as the photon traveling towards the star, which resulted in miscalculation of the interaction angle at every step of the integral along the line-of-sight and eventually in overestimation of the optical depth both for the stellar radiation and the circumstellar disc (Fig.~\ref{fig:old_new}). This mistake does not considerably change the shape of the orbital evolution of the optical depth. Qualitatively the results presented in \citet{2017ApJ...837..175S} are not strongly affected, but the overall effect of the gamma-gamma absorption is expected to be weaker. All the results will be updated in the forthcoming dedicated paper on PSR~B1259-63/LS 2883. 

\section{Comparison of analytic and numeric approach}
\label{a:anVSnum}

\begin{figure*}
\centering  
  \includegraphics[width=\hsize]{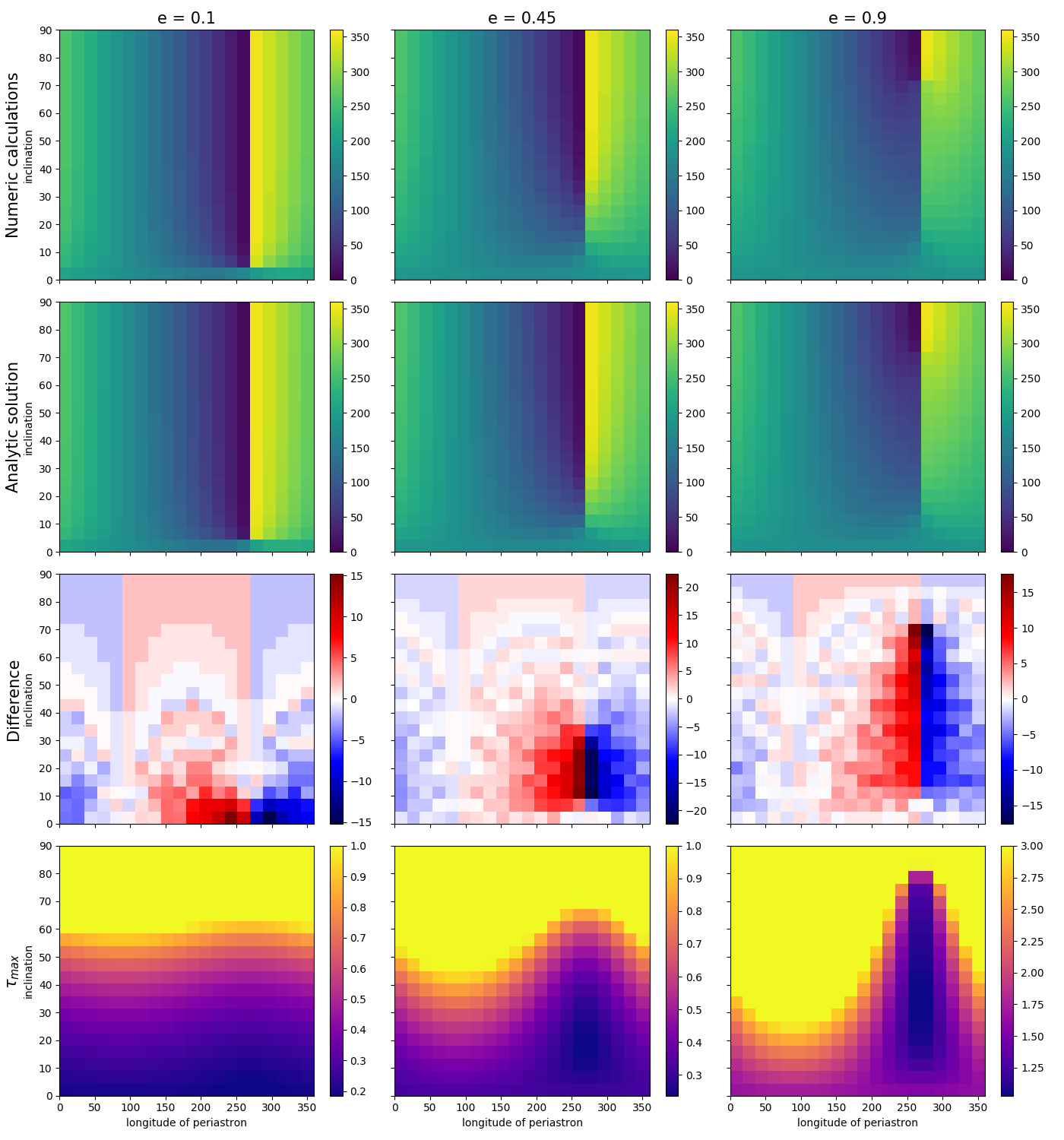}
  \caption{The upper two rows show the maps of the orbital phase of maximum absorption, $phi_\mathrm{max}$, color-coded from $0^\circ$ to $360^\circ$, calculated numerically (top panels) and analytically (upper middle panels) for the Test Binary for three different values of eccentricity and the orbital period of 100 days. The lower middle row shows the mapped difference, $\phi_\mathrm{max}^\mathrm{num} - \phi_\mathrm{max}^\mathrm{an}$, and the bottom maps show numerically calculated maximum optical depth, $\tau_{\gamma\gamma}(\phi_\mathrm{max}^\mathrm{num})$.}
  \label{fig:phiMAX_ecc}
\end{figure*}

\begin{figure*}
\centering  
  \includegraphics[width=\hsize]{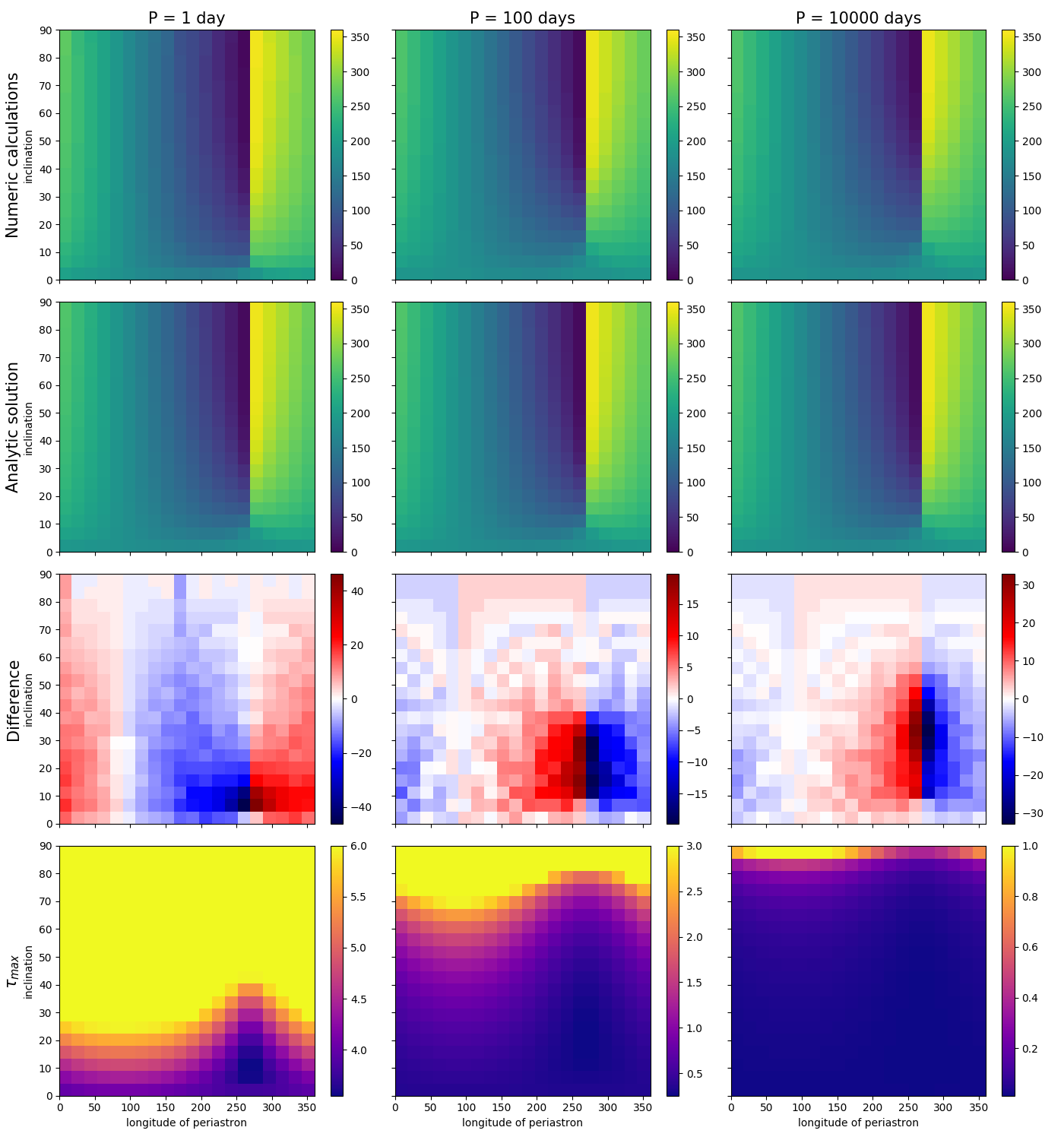}
  \caption{Same as Fig.~\ref{fig:phiMAX_ecc} but for different values of the orbital period and $e = 0.5$}
  \label{fig:phiMAX_compactness}
\end{figure*}

To calculate the orbital phase with the maximum absorption, $\phi_\mathrm{max}$, we split the orbit into 200 bins setting a systematic error of $1.8^\circ$. We estimate $\phi_\mathrm{max}$ for different combinations of $i$ and $\omega$ varying the inclination between 0 and 90 degrees (20 bins) and longitude of periastron between 0 and 360 degrees (20 bins). Figures~\ref{fig:phiMAX_ecc} and \ref{fig:phiMAX_compactness} show maps depicting color-coded $\phi_\mathrm{max}$ as a function of $i$ (y-axis) and $\omega$ (x-axis) calculated numerically (top panels) and analytically (upper middle panels). The lower middle panels show the difference between two estimates, $\phi_\mathrm{max}^\mathrm{num} - \phi_\mathrm{max}^\mathrm{an}$, and the bottom panels map the the maximum optical depth as obtained from numerical simulations, i.e. the optical depth at $\phi_\mathrm{max}^\mathrm{num}$. The stellar parameters used for these simulations correspond to the Test Binary entry in the Table~\ref{tab:binaries}. Figure~\ref{fig:phiMAX_ecc} explores the dependence on the eccentricity showing results for $e = 0.1$ (left column), $e = 0.45$ (middle column), and $e = 0.9$ (right column) for the same orbital period of 100 year, while Figure~\ref{fig:phiMAX_compactness} shows the dependence on the period for the same eccentricity of $e = 0.5$ with $P = 10$~days (left column), $P = 100$~days (middle column), and $P = 10,000$~days (right column).


As expected $\phi_\mathrm{max}$ strongly depends on the longitude of periastron of the system as long as the system is sufficiently inclined, while for low inclination angles (observer looks at the system face-on) $\omega$ is irrelevant for the estimate of $\phi_\mathrm{max}$ as it stays at $\approx 180^\circ$ (periastron) determined by the \add{to the smallest binary separation at this phase.}
For $i\gtrsim10^\circ$,  $\phi_\mathrm{max}$ does not strongly depend on $i$ for the orbits with low eccentricity and is mainly determined by the longitude of periastron
For high eccentricities the dependence on the inclination angle becomes more significant in the domain where $\omega$ is close to $270^\circ$ (periastron coincides with inferior conjunction). This is an exclusively geometrical effect reflecting the fact that for high eccentricity orbits the emitted gamma-ray photon would be traveling closer to the star when emitted at some offset to the apastron and hence the maximum absorption would occur at some orbital phase either before or after the periastron. Additionally, as mentioned previously for high eccentricity orbits the optical depth curve for $\omega\sim270^\circ$ features two maxima that further complicates things. 

Apart for a small domain around $\omega \simeq 270^\circ$ and the case with a very low orbital period, numeric and analytic results are in good agreement with the difference at the level of the systematical error of the numerical calculations. 
In the domain around $\omega \simeq 270^\circ$ the analytic solution becomes significantly different (although still at most by $\sim 15^\circ$) and it becomes more relevant at high eccentricities as this uncertainty region covers a larger range of inclination angles. However, this part of the parameter space also corresponds to weaker absorption as can be seen from the bottom panels which maps $\tau_{\gamma\gamma}(\phi_\mathrm{max}^\mathrm{num})$. 
Indeed, in this case the difference between the numeric and analytic solution can be irrelevant as the absorption \add{might simply be not} strong enough to modify the observed light curve and hence the proposed method cannot be applied. Note, that for the high eccentricity case (right column in Fig.~\ref{fig:phiMAX_ecc}) although the the difference between the numeric and analytic solutions does correlate with the weakness of the absorption, the optical depth is high for the whole parameter space. This is due to the chosen parameters as for the $e=0.9$ \add{case} the orbital period of 100 days corresponds to a very small separation distance between the pulsar and the star at periastron. These compact binary systems, however, are also irrelevant to the proposed method simply because most of the gamma-ray emission will be absorbed at any orbital phase and therefore not detected.

The numeric estimate of $\phi_\mathrm{max}$ is not very sensitive to the orbital period as long as it is not too short. There is no obvious difference between the maps for $P = 100$ days and $P=10000$ days, which is confirmed by almost identical ``difference" maps. Note, all three maps depicting the analytic solutions are identical as the analytical solution is independent of the orbital period (see Section~\ref{sec:analytic}). For very short periods, i.e. compact, systems (see $P = 10$ days column) the difference becomes more apparent, mainly due to a non-negligible size of the star. Nevertheless, as already mentioned above, these very compact systems are not of interest to us due to the impossibility of their detection.   

\bibliography{bibliography}{}

\begin{thebibliography}{}
\expandafter\ifx\csname natexlab\endcsname\relax\def\natexlab#1{#1}\fi
\providecommand{\url}[1]{\href{#1}{#1}}
\providecommand{\dodoi}[1]{doi:~\href{http://doi.org/#1}{\nolinkurl{#1}}}
\providecommand{\doeprint}[1]{\href{http://ascl.net/#1}{\nolinkurl{http://ascl.net/#1}}}
\providecommand{\doarXiv}[1]{\href{https://arxiv.org/abs/#1}{\nolinkurl{https://arxiv.org/abs/#1}}}

\bibitem[{{Abdo} {et~al.}(2009){Abdo}, {Ackermann}, {Ajello}, {Anderson}, {Atwood}, {Axelsson}, {Baldini}, {Ballet}, {Barbiellini}, {Baring}, {Bastieri}, {Baughman}, {Bechtol}, {Bellazzini}, {Berenji}, {Bignami}, {Blandford}, {Bloom}, {Bonamente}, {Borgland}, {Bregeon}, {Brez}, {Brigida}, {Bruel}, {Burnett}, {Caliandro}, {Cameron}, {Caraveo}, {Casandjian}, {Cecchi}, {{\c{C}}elik}, {Chekhtman}, {Cheung}, {Chiang}, {Ciprini}, {Claus}, {Cohen-Tanugi}, {Conrad}, {Cutini}, {Dermer}, {de Angelis}, {de Luca}, {de Palma}, {Digel}, {Dormody}, {do Couto e Silva}, {Drell}, {Dubois}, {Dumora}, {Farnier}, {Favuzzi}, {Fegan}, {Fukazawa}, {Funk}, {Fusco}, {Gargano}, {Gasparrini}, {Gehrels}, {Germani}, {Giebels}, {Giglietto}, {Giommi}, {Giordano}, {Glanzman}, {Godfrey}, {Grenier}, {Grondin}, {Grove}, {Guillemot}, {Guiriec}, {Gwon}, {Hanabata}, {Harding}, {Hayashida}, {Hays}, {Hughes}, {J{\'o}hannesson}, {Johnson}, {Johnson}, {Johnson}, {Kamae}, {Katagiri}, {Kataoka}, {Kawai}, {Kerr}, {Kn{\"o}dlseder}, {Kocian}, {Kuss},
  {Lande}, {Latronico}, {Lemoine-Goumard}, {Longo}, {Loparco}, {Lott}, {Lovellette}, {Lubrano}, {Madejski}, {Makeev}, {Marelli}, {Mazziotta}, {McConville}, {McEnery}, {Meurer}, {Michelson}, {Mitthumsiri}, {Mizuno}, {Monte}, {Monzani}, {Morselli}, {Moskalenko}, {Murgia}, {Nolan}, {Norris}, {Nuss}, {Ohsugi}, {Omodei}, {Orlando}, {Ormes}, {Paneque}, {Parent}, {Pelassa}, {Pepe}, {Pesce-Rollins}, {Pierbattista}, {Piron}, {Porter}, {Primack}, {Rain{\`o}}, {Rando}, {Ray}, {Razzano}, {Rea}, {Reimer}, {Reimer}, {Reposeur}, {Ritz}, {Rochester}, {Rodriguez}, {Romani}, {Ryde}, {Sadrozinski}, {Sanchez}, {Sander}, {Parkinson}, {Scargle}, {Sgr{\`o}}, {Siskind}, {Smith}, {Smith}, {Spandre}, {Spinelli}, {Starck}, {Strickman}, {Suson}, {Tajima}, {Takahashi}, {Takahashi}, {Tanaka}, {Thayer}, {Thompson}, {Tibaldo}, {Tibolla}, {Torres}, {Tosti}, {Tramacere}, {Uchiyama}, {Usher}, {Van Etten}, {Vasileiou}, {Vilchez}, {Vitale}, {Waite}, {Wang}, {Watters}, {Winer}, {Wolff}, {Wood}, {Ylinen}, {Ziegler}, \& {Fermi LAT
  Collaboration}}]{2009Sci...325..840A}
{Abdo}, A.~A., {Ackermann}, M., {Ajello}, M., {et~al.} 2009, Science, 325, 840, \dodoi{10.1126/science.1175558}

\bibitem[{{Abeysekara} {et~al.}(2018){Abeysekara}, {Benbow}, {Bird}, {Brill}, {Brose}, {Buckley}, {Chromey}, {Daniel}, {Falcone}, {Finley}, \& et~al.}]{2018ApJ...867L..19A}
{Abeysekara}, A.~U., {Benbow}, W., {Bird}, R., {et~al.} 2018, \apjl, 867, L19, \dodoi{10.3847/2041-8213/aae70e}

\bibitem[{{Aharonian} {et~al.}(2005){Aharonian}, {Akhperjanian}, {Aye}, {Bazer-Bachi}, {Beilicke}, {Benbow}, {Berge}, {Berghaus}, {Bernl{\"o}hr}, {Boisson}, {Bolz}, {Braun}, {Breitling}, {Brown}, {Bussons Gordo}, {Chadwick}, {Chounet}, {Cornils}, {Costamante}, {Degrange}, {Djannati-Ata{\"i}}, {O'C.~Drury}, {Dubus}, {Emmanoulopoulos}, {Espigat}, {Feinstein}, {Fleury}, {Fontaine}, {Fuchs}, {Funk}, {Gallant}, {Giebels}, {Gillessen}, {Glicenstein}, {Goret}, {Hadjichristidis}, {Hauser}, {Heinzelmann}, {Henri}, {Hermann}, {Hinton}, {Hofmann}, {Holleran}, {Horns}, {de Jager}, {Johnston}, {Kh{\'e}lifi}, {Kirk}, {Komin}, {Konopelko}, {Latham}, {Le Gallou}, {Lemi{\`e}re}, {Lemoine-Goumard}, {Leroy}, {Martineau-Huynh}, {Lohse}, {Marcowith}, {Masterson}, {McComb}, {de Naurois}, {Nolan}, {Noutsos}, {Orford}, {Osborne}, {Ouchrif}, {Panter}, {Pelletier}, {Pita}, {P{\"u}hlhofer}, {Punch}, {Raubenheimer}, {Raue}, {Raux}, {Rayner}, {Redondo}, {Reimer}, {Reimer}, {Ripken}, {Rob}, {Rolland}, {Rowell}, {Sahakian}, {Saug{\'e}},
  {Schlenker}, {Schlickeiser}, {Schuster}, {Schwanke}, {Siewert}, {Skj{\ae}raasen}, {Sol}, {Steenkamp}, {Stegmann}, {Tavernet}, {Terrier}, {Th{\'e}oret}, {Tluczykont}, {Vasileiadis}, {Venter}, {Vincent}, {V{\"o}lk}, \& {Wagner}}]{2005A&A...442....1A}
{Aharonian}, F., {Akhperjanian}, A.~G., {Aye}, K.-M., {et~al.} 2005, \aap, 442, 1, \dodoi{10.1051/0004-6361:20052983}

\bibitem[{{Aharonian} {et~al.}(2009){Aharonian}, {Akhperjanian}, {Anton}, {Barres de Almeida}, {Bazer-Bachi}, {Becherini}, {Behera}, {Bernl{\"o}hr}, {Bochow}, {Boisson}, {Bolmont}, {Borrel}, {Brucker}, {Brun}, {Brun}, {B{\"u}hler}, {Bulik}, {B{\"u}sching}, {Boutelier}, {Chadwick}, {Charbonnier}, {Chaves}, {Cheesebrough}, {Chounet}, {Clapson}, {Coignet}, {Dalton}, {Daniel}, {Davids}, {Degrange}, {Deil}, {Dickinson}, {Djannati-Ata{\"i}}, {Domainko}, {O'C.~Drury}, {Dubois}, {Dubus}, {Dyks}, {Dyrda}, {Egberts}, {Emmanoulopoulos}, {Espigat}, {Farnier}, {Feinstein}, {Fiasson}, {F{\"o}rster}, {Fontaine}, {F{\"u}{\ss}ling}, {Gabici}, {Gallant}, {G{\'e}rard}, {Gerbig}, {Giebels}, {Glicenstein}, {Gl{\"u}ck}, {Goret}, {G{\"o}ring}, {Hauser}, {Hauser}, {Heinz}, {Heinzelmann}, {Henri}, {Hermann}, {Hinton}, {Hoffmann}, {Hofmann}, {Holleran}, {Hoppe}, {Horns}, {Jacholkowska}, {de Jager}, {Jahn}, {Jung}, {Katarzy{\'n}ski}, {Katz}, {Kaufmann}, {Kerschhaggl}, {Khangulyan}, {Kh{\'e}lifi}, {Keogh}, {Klochkov}, {Klu{\'z}niak},
  {Kneiske}, {Komin}, {Kosack}, {Kossakowski}, {Lamanna}, {Lenain}, {Lohse}, {Marandon}, {Martineau-Huynh}, {Marcowith}, {Masbou}, {Maurin}, {McComb}, {Medina}, {Moderski}, {Moulin}, {Naumann-Godo}, {de Naurois}, {Nedbal}, {Nekrassov}, {Nicholas}, {Niemiec}, {Nolan}, {Ohm}, {Olive}, {de O{\~n}a Wilhelmi}, {Orford}, {Ostrowski}, {Panter}, {Paz Arribas}, {Pedaletti}, {Pelletier}, {Petrucci}, {Pita}, {P{\"u}hlhofer}, {Punch}, {Quirrenbach}, {Raubenheimer}, {Raue}, {Rayner}, {Renaud}, {Rieger}, {Ripken}, {Rob}, {Rosier-Lees}, {Rowell}, {Rudak}, {Rulten}, {Ruppel}, {Sahakian}, {Santangelo}, {Schlickeiser}, {Sch{\"o}ck}, {Schwanke}, {Schwarzburg}, {Schwemmer}, {Shalchi}, {Sikora}, {Skilton}, {Sol}, {Spangler}, {Stawarz}, {Steenkamp}, {Stegmann}, {Stinzing}, {Superina}, {Szostek}, {Tam}, {Tavernet}, {Terrier}, {Tibolla}, {Tluczykont}, {van Eldik}, {Vasileiadis}, {Venter}, {Venter}, {Vialle}, {Vincent}, {Vivier}, {V{\"o}lk}, {Volpe}, {Wagner}, {Ward}, {Zdziarski}, \& {Zech}}]{2009A&A...507..389A}
{Aharonian}, F., {Akhperjanian}, A.~G., {Anton}, G., {et~al.} 2009, \aap, 507, 389, \dodoi{10.1051/0004-6361/200912339}

\bibitem[{{Aragona} {et~al.}(2009){Aragona}, {McSwain}, {Grundstrom}, {Marsh}, {Roettenbacher}, {Hessler}, {Boyajian}, \& {Ray}}]{2009ApJ...698..514A}
{Aragona}, C., {McSwain}, M.~V., {Grundstrom}, E.~D., {et~al.} 2009, \apj, 698, 514, \dodoi{10.1088/0004-637X/698/1/514}

\bibitem[{{Boettcher} {et~al.}(2012){Boettcher}, {Harris}, \& {Krawczynski}}]{2012rjag.book.....B}
{Boettcher}, M., {Harris}, D.~E., \& {Krawczynski}, H. 2012, {Relativistic Jets from Active Galactic Nuclei} (Wiley)

\bibitem[{{Bosch-Ramon} {et~al.}(2015){Bosch-Ramon}, {Barkov}, \& {Perucho}}]{2015A&A...577A..89B}
{Bosch-Ramon}, V., {Barkov}, M.~V., \& {Perucho}, M. 2015, \aap, 577, A89, \dodoi{10.1051/0004-6361/201425228}

\bibitem[{{Bosch-Ramon} \& {Khangulyan}(2009)}]{2009IJMPD..18..347B}
{Bosch-Ramon}, V., \& {Khangulyan}, D. 2009, International Journal of Modern Physics D, 18, 347, \dodoi{10.1142/S0218271809014601}

\bibitem[{{B{\"o}ttcher} \& {Dermer}(2005)}]{2005ApJ...634L..81B}
{B{\"o}ttcher}, M., \& {Dermer}, C.~D. 2005, \apjl, 634, L81, \dodoi{10.1086/498615}

\bibitem[{{Camilo} {et~al.}(2009){Camilo}, {Ray}, {Ransom}, {Burgay}, {Johnson}, {Kerr}, {Gotthelf}, {Halpern}, {Reynolds}, {Romani}, {Demorest}, {Johnston}, {van Straten}, {Saz Parkinson}, {Ziegler}, {Dormody}, {Thompson}, {Smith}, {Harding}, {Abdo}, {Crawford}, {Freire}, {Keith}, {Kramer}, {Roberts}, {Weltevrede}, \& {Wood}}]{2009ApJ...705....1C}
{Camilo}, F., {Ray}, P.~S., {Ransom}, S.~M., {et~al.} 2009, \apj, 705, 1, \dodoi{10.1088/0004-637X/705/1/1}

\bibitem[{{Casares} {et~al.}(2005){Casares}, {Rib{\'o}}, {Ribas}, {Paredes}, {Mart{\'\i}}, \& {Herrero}}]{2005MNRAS.364..899C}
{Casares}, J., {Rib{\'o}}, M., {Ribas}, I., {et~al.} 2005, \mnras, 364, 899, \dodoi{10.1111/j.1365-2966.2005.09617.x}

\bibitem[{{Chernyakova} \& {Malyshev}(2020)}]{2020mbhe.confE..45C}
{Chernyakova}, M., \& {Malyshev}, D. 2020, in Multifrequency Behaviour of High Energy Cosmic Sources - XIII. 3-8 June 2019. Palermo, 45, \dodoi{10.22323/1.362.0045}

\bibitem[{{Chernyakova} {et~al.}(2019){Chernyakova}, {Malyshev}, {Paizis}, {La Palombara}, {Balbo}, {Walter}, {Hnatyk}, {van Soelen}, {Romano}, {Munar-Adrover}, {Vovk}, {Piano}, {Capitanio}, {Falceta-Gon{\c{c}}alves}, {Landoni}, {Luque-Escamilla}, {Mart{\'\i}}, {Paredes}, {Rib{\'o}}, {Safi-Harb}, {Saha}, {Sidoli}, \& {Vercellone}}]{cta_binary19}
{Chernyakova}, M., {Malyshev}, D., {Paizis}, A., {et~al.} 2019, \aap, 631, A177, \dodoi{10.1051/0004-6361/201936501}

\bibitem[{{Coe} {et~al.}(2019){Coe}, {Okazaki}, {Steele}, {Ng}, {Ho}, {Lyne}, {Stappers}, {Johnson}, {Ray}, \& {Kerr}}]{2019MNRAS.485.1864C}
{Coe}, M.~J., {Okazaki}, A.~T., {Steele}, I.~A., {et~al.} 2019, \mnras, 485, 1864, \dodoi{10.1093/mnras/stz515}

\bibitem[{{Dermer} \& {B{\"o}ttcher}(2006)}]{2006ApJ...643.1081D}
{Dermer}, C.~D., \& {B{\"o}ttcher}, M. 2006, \apj, 643, 1081, \dodoi{10.1086/502966}

\bibitem[{{Dubus}(2006{\natexlab{a}})}]{2006A&A...456..801D}
{Dubus}, G. 2006{\natexlab{a}}, \aap, 456, 801, \dodoi{10.1051/0004-6361:20054779}

\bibitem[{{Dubus}(2006{\natexlab{b}})}]{2006A&A...451....9D}
---. 2006{\natexlab{b}}, \aap, 451, 9, \dodoi{10.1051/0004-6361:20054233}

\bibitem[{{Dubus}(2013)}]{2013A&ARv..21...64D}
---. 2013, \aapr, 21, 64, \dodoi{10.1007/s00159-013-0064-5}

\bibitem[{{Dubus}(2015)}]{2015CRPhy..16..661D}
---. 2015, Comptes Rendus Physique, 16, 661, \dodoi{10.1016/j.crhy.2015.08.014}

\bibitem[{{Dubus} {et~al.}(2015){Dubus}, {Lamberts}, \& {Fromang}}]{2015A&A...581A..27D}
{Dubus}, G., {Lamberts}, A., \& {Fromang}, S. 2015, \aap, 581, A27, \dodoi{10.1051/0004-6361/201425394}

\bibitem[{{Gould} \& {Schr{\'e}der}(1967)}]{1967PhRv..155.1404G}
{Gould}, R.~J., \& {Schr{\'e}der}, G.~P. 1967, Physical Review, 155, 1404, \dodoi{10.1103/PhysRev.155.1404}

\bibitem[{{H.~E.~S.~S. Collaboration} {et~al.}(2020){H.~E.~S.~S. Collaboration}, {Abdalla}, {Adam}, {Aharonian}, {Ait Benkhali}, {Ang{\"u}ner}, {Arakawa}, {Arcaro}, {Armand}, {Ashkar}, {Backes}, {Barbosa Martins}, {Barnard}, {Becherini}, {Berge}, {Bernl{\"o}hr}, {Blackwell}, {B{\"o}ttcher}, {Boisson}, {Bolmont}, {Bonnefoy}, {Bregeon}, {Breuhaus}, {Brun}, {Brun}, {Bryan}, {B{\"u}chele}, {Bulik}, {Bylund}, {Caroff}, {Carosi}, {Casanova}, {Cerruti}, {Chand}, {Chandra}, {Chaves}, {Chen}, {Colafrancesco}, {Cury{\l}o}, {Davids}, {Deil}, {Devin}, {deWilt}, {Dirson}, {Djannati-Ata{\"\i}}, {Dmytriiev}, {Donath}, {Doroshenko}, {Dyks}, {Egberts}, {Emery}, {Ernenwein}, {Eschbach}, {Feijen}, {Fegan}, {Fiasson}, {Fontaine}, {Funk}, {F{\"u}{\ss}ling}, {Gabici}, {Gallant}, {Gat{\'e}}, {Giavitto}, {Giunti}, {Glawion}, {Glicenstein}, {Gottschall}, {Grondin}, {Hahn}, {Haupt}, {Heinzelmann}, {Henri}, {Hermann}, {Hinton}, {Hofmann}, {Hoischen}, {Holch}, {Holler}, {Horns}, {Huber}, {Iwasaki}, {Jamrozy}, {Jankowsky}, {Jankowsky},
  {Jardin-Blicq}, {Jung-Richardt}, {Kastendieck}, {Katarzy{\'n}ski}, {Katsuragawa}, {Katz}, {Khangulyan}, {Kh{\'e}lifi}, {King}, {Klepser}, {Klu{\'z}niak}, {Komin}, {Kosack}, {Kostunin}, {Kreter}, {Lamanna}, {Lemi{\`e}re}, {Lemoine-Goumard}, {Lenain}, {Leser}, {Levy}, {Lohse}, {Lypova}, {Mackey}, {Majumdar}, {Malyshev}, {Malyshev}, {Marandon}, {Marcowith}, {Mares}, {Mariaud}, {Mart{\'\i}-Devesa}, {Marx}, {Maurin}, {Meintjes}, {Mitchell}, {Moderski}, {Mohamed}, {Mohrmann}, {Moore}, {Moulin}, {Muller}, {Murach}, {Nakashima}, {de Naurois}, {Ndiyavala}, {Niederwanger}, {Niemiec}, {Oakes}, {O'Brien}, {Odaka}, {Ohm}, {de Ona Wilhelmi}, {Ostrowski}, {Oya}, {Panter}, {Parsons}, {Perennes}, {Petrucci}, {Peyaud}, {Piel}, {Pita}, {Poireau}, {Priyana Noel}, {Prokhorov}, {Prokoph}, {P{\"u}hlhofer}, {Punch}, {Quirrenbach}, {Raab}, {Rauth}, {Reimer}, {Reimer}, {Remy}, {Renaud}, {Rieger}, {Rinchiuso}, {Romoli}, {Rowell}, {Rudak}, {Ruiz-Velasco}, {Sahakian}, {Sailer}, {Saito}, {Sanchez}, {Santangelo}, {Sasaki},
  {Schlickeiser}, {Sch{\"u}ssler}, {Schulz}, {Schutte}, {Schwanke}, {Schwemmer}, {Seglar-Arroyo}, {Senniappan}, {Seyffert}, {Shafi}, {Shiningayamwe}, {Simoni}, {Sinha}, {Sol}, {Specovius}, {Spir-Jacob}, {Stawarz}, {Steenkamp}, {Stegmann}, {Steppa}, {Takahashi}, {Tavernier}, {Taylor}, {Terrier}, {Tiziani}, {Tluczykont}, {Trichard}, {Tsirou}, {Tsuji}, {Tuffs}, {Uchiyama}, {van der Walt}, {van Eldik}, {van Rensburg}, {van Soelen}, {Vasileiadis}, {Veh}, {Venter}, {Vincent}, {Vink}, {V{\"o}lk}, {Vuillaume}, {Wadiasingh}, {Wagner}, {White}, {Wierzcholska}, {Yang}, {Yoneda}, {Zacharias}, {Zanin}, {Zdziarski}, {Zech}, {Zorn}, {{\.Z}ywucka}, \& {Bordas}}]{2020A&A...633A.102H}
{H.~E.~S.~S. Collaboration}, {Abdalla}, H., {Adam}, R., {et~al.} 2020, \aap, 633, A102, \dodoi{10.1051/0004-6361/201936621}

\bibitem[{{H.E.S.S.~Collaboration} {et~al.}(2013){H.E.S.S.~Collaboration}, {Abramowski}, {Acero}, {Aharonian}, {Akhperjanian}, {Anton}, {Balenderan}, {Balzer}, {Barnacka}, {Becherini}, {Becker Tjus}, {Bernl{\"o}hr}, {Birsin}, {Biteau}, {Boisson}, {Bolmont}, {Bordas}, {Brucker}, {Brun}, {Brun}, {Bulik}, {Carrigan}, {Casanova}, {Cerruti}, {Chadwick}, {Chaves}, {Cheesebrough}, {Colafrancesco}, {Cologna}, {Conrad}, {Couturier}, {Dalton}, {Daniel}, {Davids}, {Degrange}, {Deil}, {deWilt}, {Dickinson}, {Djannati-Ata{\"i}}, {Domainko}, {Drury}, {Dubus}, {Dutson}, {Dyks}, {Dyrda}, {Egberts}, {Eger}, {Espigat}, {Fallon}, {Farnier}, {Fegan}, {Feinstein}, {Fernandes}, {Fernandez}, {Fiasson}, {Fontaine}, {F{\"o}rster}, {F{\"u}{\ss}ling}, {Gajdus}, {Gallant}, {Garrigoux}, {Gast}, {Giebels}, {Glicenstein}, {Gl{\"u}ck}, {G{\"o}ring}, {Grondin}, {Grudzi{\'n}ska}, {H{\"a}er}, {Hague}, {Hahn}, {Hampf}, {Harris}, {Heinz}, {Heinzelmann}, {Henri}, {Hermann}, {Hillert}, {Hinton}, {Hofmann}, {Hofverberg}, {Holler}, {Horns},
  {Jacholkowska}, {Jahn}, {Jamrozy}, {Jung}, {Kastendieck}, {Katarzy{\'n}ski}, {Katz}, {Kaufmann}, {Kh{\'e}lifi}, {Klepser}, {Klochkov}, {Klu{\'z}niak}, {Kneiske}, {Kolitzus}, {Komin}, {Kosack}, {Kossakowski}, {Krayzel}, {Kr{\"u}ger}, {Lan}, {Lamanna}, {Lefaucheur}, {Lemoine-Goumard}, {Lenain}, {Lennarz}, {Lohse}, {Lopatin}, {Lu}, {Marandon}, {Marcowith}, {Masbou}, {Maurin}, {Maxted}, {Mayer}, {McComb}, {Medina}, {M{\'e}hault}, {Menzler}, {Moderski}, {Mohamed}, {Moulin}, {Naumann}, {Naumann-Godo}, {de Naurois}, {Nedbal}, {Nguyen}, {Niemiec}, {Nolan}, {Oakes}, {Ohm}, {de O{\~n}a Wilhelmi}, {Opitz}, {Ostrowski}, {Oya}, {Panter}, {Parsons}, {Paz Arribas}, {Pekeur}, {Pelletier}, {Perez}, {Petrucci}, {Peyaud}, {Pita}, {P{\"u}hlhofer}, {Punch}, {Quirrenbach}, {Raab}, {Raue}, {Reimer}, {Reimer}, {Renaud}, {de los Reyes}, {Rieger}, {Ripken}, {Rob}, {Rosier-Lees}, {Rowell}, {Rudak}, {Rulten}, {Sahakian}, {Sanchez}, {Santangelo}, {Schlickeiser}, {Schulz}, {Schwanke}, {Schwarzburg}, {Schwemmer}, {Sheidaei}, {Skilton},
  {Sol}, {Spengler}, {Stawarz}, {Steenkamp}, {Stegmann}, {Stinzing}, {Stycz}, {Sushch}, {Szostek}, {Tavernet}, {Terrier}, {Tluczykont}, {Trichard}, {Valerius}, {van Eldik}, {Vasileiadis}, {Venter}, {Viana}, {Vincent}, {V{\"o}lk}, {Volpe}, {Vorobiov}, {Vorster}, {Wagner}, {Ward}, {White}, {Wierzcholska}, {Willmann}, {Wouters}, {Zacharias}, {Zajczyk}, {Zdziarski}, {Zech}, \& {Zechlin}}]{2013A&A...551A..94H}
{H.E.S.S.~Collaboration}, {Abramowski}, A., {Acero}, F., {et~al.} 2013, \aap, 551, A94, \dodoi{10.1051/0004-6361/201220612}

\bibitem[{{Ho} {et~al.}(2017){Ho}, {Ng}, {Lyne}, {Stappers}, {Coe}, {Halpern}, {Johnson}, \& {Steele}}]{2017MNRAS.464.1211H}
{Ho}, W.~C.~G., {Ng}, C.-Y., {Lyne}, A.~G., {et~al.} 2017, MNRAS, 464, 1211, \dodoi{10.1093/mnras/stw2420}

\bibitem[{{Huber} {et~al.}(2021{\natexlab{a}}){Huber}, {Kissmann}, {Reimer}, \& {Reimer}}]{2021A&A...646A..91H}
{Huber}, D., {Kissmann}, R., {Reimer}, A., \& {Reimer}, O. 2021{\natexlab{a}}, \aap, 646, A91, \dodoi{10.1051/0004-6361/202039277}

\bibitem[{{Huber} {et~al.}(2021{\natexlab{b}}){Huber}, {Kissmann}, \& {Reimer}}]{2021A&A...649A..71H}
{Huber}, D., {Kissmann}, R., \& {Reimer}, O. 2021{\natexlab{b}}, \aap, 649, A71, \dodoi{10.1051/0004-6361/202039278}

\bibitem[{{Jauch} \& {Rohrlich}(1976)}]{1976tper.book.....J}
{Jauch}, J.~M., \& {Rohrlich}, F. 1976, {The theory of photons and electrons. The relativistic quantum field theory of charged particles with spin one-half}

\bibitem[{{Khangulyan} {et~al.}(2008){Khangulyan}, {Aharonian}, \& {Bosch-Ramon}}]{2008MNRAS.383..467K}
{Khangulyan}, D., {Aharonian}, F., \& {Bosch-Ramon}, V. 2008, \mnras, 383, 467, \dodoi{10.1111/j.1365-2966.2007.12572.x}

\bibitem[{{Lyne} {et~al.}(2015){Lyne}, {Stappers}, {Keith}, {Ray}, {Kerr}, {Camilo}, \& {Johnson}}]{2015MNRAS.451..581L}
{Lyne}, A.~G., {Stappers}, B.~W., {Keith}, M.~J., {et~al.} 2015, \mnras, 451, 581, \dodoi{10.1093/mnras/stv236}

\bibitem[{{Negueruela} {et~al.}(2011){Negueruela}, {Rib{\'o}}, {Herrero}, {Lorenzo}, {Khangulyan}, \& {Aharonian}}]{2011ApJ...732L..11N}
{Negueruela}, I., {Rib{\'o}}, M., {Herrero}, A., {et~al.} 2011, \apjl, 732, L11, \dodoi{10.1088/2041-8205/732/1/L11}

\bibitem[{{Ng} {et~al.}(2019){Ng}, {Ho}, {Gotthelf}, {Halpern}, {Coe}, {Stappers}, {Lyne}, {Wood}, \& {Kerr}}]{2019ApJ...880..147N}
{Ng}, C.~Y., {Ho}, W.~C.~G., {Gotthelf}, E.~V., {et~al.} 2019, \apj, 880, 147, \dodoi{10.3847/1538-4357/ab2adb}

\bibitem[{Olivera~Nieto(2023)}]{Laura:ss433}
Olivera~Nieto, L. 2023, PhD thesis, MPIK, Heidelberg, \dodoi{10.11588/heidok.00032936}

\bibitem[{{Romoli} {et~al.}(2015){Romoli}, {Bordas}, {Mariaud}, {Murach}, {Aharonian}, {de Naurois}, {P{\"u}hlhofer}, {Schwanke}, {van Soelen}, {Sushch}, {Zabalza}, \& {for the H.~E.~S.~S.~Collaboration}}]{2015arXiv150903090R}
{Romoli}, C., {Bordas}, P., {Mariaud}, C., {et~al.} 2015, ArXiv e-prints.
\newblock \doarXiv{1509.03090}

\bibitem[{{Sushch} \& {van Soelen}(2017)}]{2017ApJ...837..175S}
{Sushch}, I., \& {van Soelen}, B. 2017, \apj, 837, 175, \dodoi{10.3847/1538-4357/aa62ff}

\bibitem[{{Weng} {et~al.}(2022){Weng}, {Qian}, {Wang}, {Torres}, {Papitto}, {Jiang}, {Xu}, {Li}, {Yan}, {Liu}, {Ge}, \& {Yuan}}]{2022NatAs...6..698W}
{Weng}, S.-S., {Qian}, L., {Wang}, B.-J., {et~al.} 2022, Nature Astronomy, 6, 698, \dodoi{10.1038/s41550-022-01630-1}

\bibitem[{{Zdziarski} \& {Lightman}(1985)}]{1985ApJ...294L..79Z}
{Zdziarski}, A.~A., \& {Lightman}, A.~P. 1985, \apjl, 294, L79, \dodoi{10.1086/184513}

\end{thebibliography}
\bibliographystyle{aasjournal}



\end{document}